\documentclass{emulateapj}

\usepackage{natbib}

\usepackage{graphicx}
\usepackage{latexsym}
\usepackage{amsfonts,amsmath,amssymb}
\usepackage{url}
\usepackage[utf8]{inputenc}
\usepackage{fancyref}
\usepackage{hyperref}

\begin{document}

\title{TWO-DIMENSIONAL NUMERICAL SIMULATIONS OF SUPERCRITICAL ACCRETION FLOWS REVISITED }

\author{Xiao-Hong Yang\altaffilmark{1,}\altaffilmark{2}, Feng
Yuan\altaffilmark{1}, Ken Ohsuga\altaffilmark{3}, and De-Fu
Bu\altaffilmark{1}}

\altaffiltext{1}{Shanghai Astronomical Observatory, Chinese Academy
of Sciences, 80 Nandan Road, Shanghai 200030, China;
yangxh@cqu.edu.cn, fyuan@shao.ac.cn} \altaffiltext{2}{Department of
Physics, Chongqing University, Chongqing 400044, China}
\altaffiltext{3}{National Astronomical Observatory of Japan, Osawa,
Mitaka, Tokyo 181-8588, Japan}


\begin{abstract}
We study the dynamics of super-Eddington accretion flows by
performing two-dimensional radiation-hydrodynamic simulations.
Compared with previous works, in this paper we include the
$T_{\theta\phi}$ component of the viscous stress and consider
various values of the viscous parameter $\alpha$. We find that when
$T_{\theta\phi}$ is included, the rotational speed of the
high-latitude flow decreases, while the density increases and
decreases at the high and low latitudes, respectively. We calculate
the radial profiles of inflow and outflow rates. We find that the
inflow rate decreases inward, following a power law form of
$\dot{M}_{\rm in}\propto r^s$. The value of $s$ depends on the
magnitude of $\alpha$ and is within the range of $\sim 0.4-1.0$.
Correspondingly, the radial profile of density becomes flatter
compared with the case of a constant $\dot{M}(r)$. We find that the
density profile can be described by $\rho(r)\propto r^{-p}$ and the
value of $p$ is almost same for a wide range of $\alpha$ ranging
from $\alpha=0.1$ to $0.005$. The inward decrease of inflow
accretion rate is very similar to hot accretion flows, which is
attributed to the mass loss in outflows. To study the origin of
outflow, we analyze the convective stability of the slim disk. We find
that depending on the value of $\alpha$, the flow is marginally
stable (when $\alpha$ is small) or unstable (when $\alpha$ is
large). This is different from the case of hydrodynamical hot
accretion flow where radiation is dynamically unimportant and the
flow is always convectively unstable. We speculate that the reason
for the difference is because radiation can stabilize convection.
The origin of outflow is thus likely because of the joint function
of convection and radiation, but further investigation is required.
\end{abstract}

\keywords{accretion, accretion disk -- black hole physics -- hydrodynamics -- methods: numerical -- radiative transfer}

\section{Introduction}

One milestone in black hole accretion is the standard thin disk
model (Shakura \& Sunyaev 1973; Pringle 1981). This model applies
below the Eddington accretion rate defined as $\dot{M}_{\rm
Edd}\equiv 10L_{\rm Edd}/c^2$ ($L_{\rm Edd}$ is the Eddington
luminosity). In this model, all the viscously dissipated energy is
immediately radiated away. When the accretion rate is above the
Eddington rate, advection begins to become important and the
accretion model is described by the ``slim disk'' (Abramowicz et al.
1988; see also Begelman \& Meier 1982). In a slim disk, the
radiative efficiency is lower than that of the standard thin disk
because of energy advection and photons trapping effects. The energy
dissipated in the disk is advected with the accreting matter, since
the radiative diffusion timescale is longer than the accretion
timescale. Note that the slim-disk model cannot correctly treat the
photon trapping, because photon trapping is basically a
multi-dimensional effect (Ohsuga et al. 2002; 2003). The potential
applications of slim-disks include narrow-line Seyfert galaxies
(Mineshige et al. 2000) and ultraluminous X-ray sources (Watarai et
al. 2001; Vierdayanti et al. 2006).

The above-mentioned pioneer works on slim disk are all one
dimensional and analytical. Multi-dimensional and time-dependent
numerical simulations obviously can reveal important additional
information about the dynamics of the accretion flow. Many
radiation-hydrodynamic (RHD; Eggum et al. 1987, 1988; Okuda 2002;
Okuda et al. 2005; Ohsuga et al. 2005) and radiation
magnetohydrodynamic (MHD; Ohsuga et al. 2009; Ohsuga \& Mineshige
2011) numerical simulation works on slim disk have been performed.
Among these works, Ohsuga et al. (2005) obtain a quasi-steady
structure of the supercritical accretion flows and outflows by a
two-dimensional global simulation. Their results broadly confirm the
main properties predicted by the analytical slim-disk model.
Moreover, they show that the accretion flow is composed of the disk
region around the equatorial plane and the outflow region above and
below the disk. Ohsuga \& Mineshige (2007) further identify that the
supercritical accretion is feasible because a very large radiation
energy density actually produces a small radiative flux as well as a
force, because of the large optical depth and photon trapping
effects.

Recently, there are some works studying the thermal stability of
radiation pressure dominated thin disk using shearing box MHD
numerical simulation with radiative transfer (Hirose et al. 2009;
Jiang et al. 2013). Hirose et al.(2009) found that the disk is
thermally stable, while Jiang et al. (2013) found that it is
unstable. The reason for the discrepancy is discussed in the latter
work.

In the RHD simulation of Ohsuga et al. (2005), an anomalous shear
stress is included to mimic the angular momentum transfer. However,
in reality, we expect Maxwell stresses associated with MHD
turbulence driven by the magneto-rotational instability (MRI) to
provide angular momentum transport in accretion flows (Balbus \&
Hawley 1998). Since MRI is driven only by shear associated with the
orbital dynamics, when an anomalous shear stress is adopted we
should only set the two azimuthal components ($r\phi$ and
$\theta\phi$) of the stress tensor to be non-zero (Stone et
al.1999). This is confirmed by the local shearing box simulations,
which indicate that the azimuthal components of the Maxwell stress
are one order of magnitude larger than the poloidal components
(e.g., Hawley et al.1995, 1996; Stone et al. 1996). In Ohsuga et al.
(2005) only the $r\phi$ component of the viscous stress is included,
while the $\theta\phi$ component is neglected. So our first aim in
this paper is to examine the effect of including the $\theta\phi$
component on the dynamics of the slim disk. It should play  an
important role in the angular momentum transport between different
latitudes and might suppress the Kelvin--Helmholtz instability on
the boundary between the outflow regions and the disk.

Another aim is to study the effect of the magnitude of $\alpha$.
Ohsuga et al.(2005) only consider one value of $\alpha$. Previous HD
numerical simulations of non-radiative accretion flows have shown
that the structure and dynamics of accretion flow depend on the
value of $\alpha$ (Igumenshchev \& Abramowicz 1999; Stone et al.
1999; Yuan et al. 2012). The radial profile of inflow rate becomes
significantly flatter when $\alpha$ becomes larger (Yuan et al.
2012). Different than the non-radiative accretion flow, a slim disk
is dominated by radiation. It is thus unclear whether or how the
dynamics of accretion flow depend on the value of $\alpha$.

Yet another aim of the present work, which is perhaps more
interesting, is to examine the radial profiles of inflow rate. In
almost all analytical models of accretion disks, the mass accretion
rate is assumed to be constant with radius. The validity of this
assumption, however, has never been proved. For the hot accretion
flows, both HD and MHD numerical simulations have found that the
mass inflow rate (refer to Equation (4) for definition) decreases
inward (e.g., Stone et al. 1999; Stone \& Pringle 2001; Yuan et al.
2012, and references therein). This is one of the most important
findings of global simulation of accretion flows, because this
result supplies an important clue to revealing the dynamics of
accretion flow. Various models have been proposed to explain this
result such as adiabatic inflow--outflow solution (Blandford \&
Begelman 1999, 2004; Begelman 2012) and convection-dominated
accretion flow (CDAF; Narayan et al. 2000; Quataert \& Gruzinov
2000). In the former, it is {\it assumed} that the inward decrease
of inflow rate is because of mass loss in the outflow, while in the
latter the flow is assumed to be convectively unstable, and a
convective envelope solution is constructed that can also explain
the simulations. No outflow is needed in CDAFs. Recent numerical
simulations have shown that MHD accretion flows are convectively
stable (Narayan et al. 2012; Yuan et al. 2012). Moreover, by
comparing the properties of inflow and outflow on the base of their
HD and MHD numerical simulation data, Yuan et al. (2012) argue that
mass outflow should be significant. They propose that  inward
decrease of accretion rate is due to outflow. Li et al. (2013)
obtained a similar conclusion. In most of the published papers, the
inflow and outflow rates (Equation(4) and (5)) are computed at each
instant of time by using instantaneous velocities, and then the
time-averaged. Narayan et al. (2012; see also Sadowski et al. 2013
for the case of spin black holes), however, found that the outflow
rate is much lower since they think the outflow rate should be
calculated by doing the time-average first (see Yuan et al. 2012 for
more discussions on the discrepancy of the two approaches of
calculating the outflow rate). While it is agreed that outflow
exists, so far the strength of outflow is still an open question.
F.Yuan et al. (2013, in preparation) have also studied the mechanism
of producing outflow. It was found that in the HD case it is
buoyancy associated with the convective instability, while it is
mainly magnetic centrifugal force in the case of MHD accretion flow.
It is obviously interesting to see whether the radial profile of the
inflow rate of a slim disk behaves like a hot accretion flow,and if
it does, whether the outflow is produced by the buoyancy associated
with the convective instability.

\begin{table*}[]
  \caption[]{Summary of Simulations}
  \label{}
  \begin{center}
  \begin{tabular}{ccccccccccccccccc}
  \hline\noalign{\smallskip}

Models & Run&$\alpha$ &$\dot{m}_{\rm input}$ & $\theta$ & Stress Tensor & $N_r \times N_{\theta}$ & t$_s$(orbits)&t$_f$ (orbits) &$\dot{m}_{\rm acc}$ &L\\

\hline\noalign{\smallskip}
A   &1a & 0.1   & 1000  &0$\sim$ $ \pi$ & $T_{r\phi}$ &96$\times$225&11.5&46.1&184.5&2.4\\
\hline\noalign{\smallskip}
B  &1b & 0.1   & 1000  &0$\sim$ $\pi$ & $T_{r\phi}$,$T_{\theta\phi}$&96$\times$225 &11.5&46.1& 160.0 & 2.2\\
   &2b & 0.1   & 3000  &0$\sim$ $\pi$ & $T_{r\phi}$,$T_{\theta\phi}$&96$\times$225 &11.5&46.1& 374.5 & 3.2\\
  &3b & 0.05  & 3000  &0$\sim$ $\pi$ & $T_{r\phi}$,$T_{\theta\phi}$&96$\times$225 &11.5&46.1 & 382.8& 3.6\\
  &4b& 0.01  & 3000  &0$\sim$ $\pi$ & $T_{r\phi}$,$T_{\theta\phi}$&96$\times$225 &63.5&98.1 &240.3 &4.2\\
  &5b& 0.005 & 3000  &0$\sim$ $\pi$ & $T_{r\phi}$,$T_{\theta\phi}$&96$\times$225 &150.0&184.6 &125.3&3.6\\
  \noalign{\smallskip}\hline
  \end{tabular}
  \end{center}
\begin{list}{}
\item\scriptsize{Note: Columns 1 and 2: the classification of our models and their
number, respectively. Column 3: the viscous parameter ($\alpha$). Column 4: the mass injection rate ($\dot{m}_{\rm{input}}$) in unit of the
critical mass accretion rate, $\dot{M}_{\rm crit}\equiv L_{\rm
Edd}/c^2$. Columns 5, 6, and 7: the computational domain in the
$\theta$ direction, the components of stress tensor, and the
quantity of grid cells. Column 8: the approximate time $t_s$ (in
units of the orbital time at $r$=100 $r_s$) after which the accretion
flows become quasi-steady. Column 9: the final time $t_f$. Columns 10
and 11: the mass accretion rate on the BH ($\dot{m}_{\rm acc}$ ) in
unit of the critical mass accretion rate, and the luminosity ($L$)
in unit of $L_{\rm Edd}$, respectively.}

\end{list}
\end{table*}

The paper is organized as follows. We describe our models and
numerical method in Section 2. The simulation results are presented in
Section 3.Section 4 is devoted to a summary and discussions.

\section{Models and numerical method}

The RHD equations are the same as those in Ohsuga et al. (2005). In
the RHD equations, the flux-limited diffusion (FLD) approximation
developed by Levermore \& Pomraning (1981) is adopted. We neglect
the self-gravity of the disk and use the pseudo-Newtonian potential
to mimic the general relativistic effects, $\psi=-GM/(r-r_{s})$
(Paczy\'{n}sky \& Wiita 1980), where $G$ is the gravitational
constant, $M$ is the mass of the black hole, $r_{s}=2GM / c^{2}$ the
Schwarzschild radius, and $r$ is the radius. We assume that the gas is
in local thermodynamic equilibrium and neglect the frequency
dependence of the opacities.

As we state in Section 1, we adopt a stress tensor $\bm{T}$ to mimic the
shear stress, which is in reality should be replaced by the magnetic
stress associated with MHD turbulence driven by the
MRI. In most cases, following Stone et al. (1999), we
assume that the only non-zero components of $\bm{T}$ are the
azimuthal components:
\begin{equation}
\bm{T}_{r\phi} =\eta r \frac{\partial}{\partial r}
\left(\frac{\upsilon_{\phi}}{r}\right),
\end{equation}
\begin{equation}
\bm{T}_{\theta\phi} =\frac{\eta
\sin(\theta)}{r}\frac{\partial}{\partial \theta}
\left(\frac{\upsilon_{\phi}}{\textit{\rm sin}(\theta)}\right).
\end{equation}
Here, the dynamical viscosity coefficient $\eta$ is described as a
function of the pressure $\eta=\alpha (p_g+\lambda
E_{0})/\Omega_{K}$, where $\Omega_{k}$ is the Keplerian angular
speed, $p_g$ is the gas pressure, $E_{0}$ is the radiation energy
density, and $\lambda$ is the flux limiter (Levermore \& Pomraning
1981). The viscous dissipative function is given by
$(\bm{T}_{r\phi}^2+\bm{T}_{\theta\phi}^2)/\eta$.

All of our models are calculated in spherical coordinates
($r$,$\theta$,$\phi$). The origin is set at a central black hole of
$M=10M_{\odot}$. The size of the computational domain is $2 r_{s}
\leq r \leq 500 r_{s}$ and $0 \leq \theta \leq \pi$ or $0 \leq
\theta \leq \pi/2$. The inner boundary ($r_{\rm in}$) must be smaller
than the sonic point of the accretion flow, which ensures that the
inner boundary conditions do not affect the simulation results.
Abramowicz et al. (1988) show that for a slim disk the location of the
sonic point depends on the accretion rate and the viscosity
parameter $\alpha$. For small $\alpha$ and a high accretion rate, the
sonic point locates in the range of $(2-3)r_s$. So we set
$r_{\rm in}=2r_s$.

The computational domain is divided into $N_r\times N_{\theta}$ grid
cells. A non-uniform grid is employed in the $r$ direction. The grid
points in the $r$ direction are equally distributed logarithmically,
i.e., $\triangle \ln r =$ constant. In the $\theta$ direction, in
order to better resolve the flow at the equator and to not lose the
resolution at the axis, we adopt the mixed grid. Twenty grids are
uniformly distributed within the $\pi/8$ from the axis, i.e.,
$\triangle \theta = \pi/160$. Other grids are distributed in the
angular range of $\pi/8\leq \theta \leq 7\pi/8$  or $\pi/8\leq
\theta \leq \pi/2$ in such a way that $\triangle
\cos(\theta)=3\pi/(4(N_{\theta}-20))$ or $3\pi/(8(N_{\theta}-20))$.
The outflow boundary condition is adopted at the inner radial
boundary, i.e. the values of physical variables in the ghost zones
are set to the values at the inner radial boundary. The outer radial
boundary condition is the same as that employed by Ohsuga et al.
(2005), who suggested that the matter, having a specific angular
momentum corresponding to the Keplerian angular momentum at
$r=100r_s$, is continuously injected into the computational domain
from the outer boundary near the equator. The injected gas
distributes within 0.05$\pi$ from the equator in the Gaussian function;
their radial velocity is set according to Equation(3.61) in Kato et al.
(1998), while their poloidal velocity is set to be zero. In the
angular direction, we employ axisymmetry relative to the axis and
reflection symmetry relative to the equator (when $0<\theta<\pi/2$),
respectively. The computational domain is initially filled with a
hot, rarefied, and optically thin atmosphere. The numerical approach
can be found in Ohsuga et al. (2005).

\begin{figure*}
\centering
\scalebox{0.78}[0.90]{\rotatebox{0}{\includegraphics[bb=61 358 494
716]{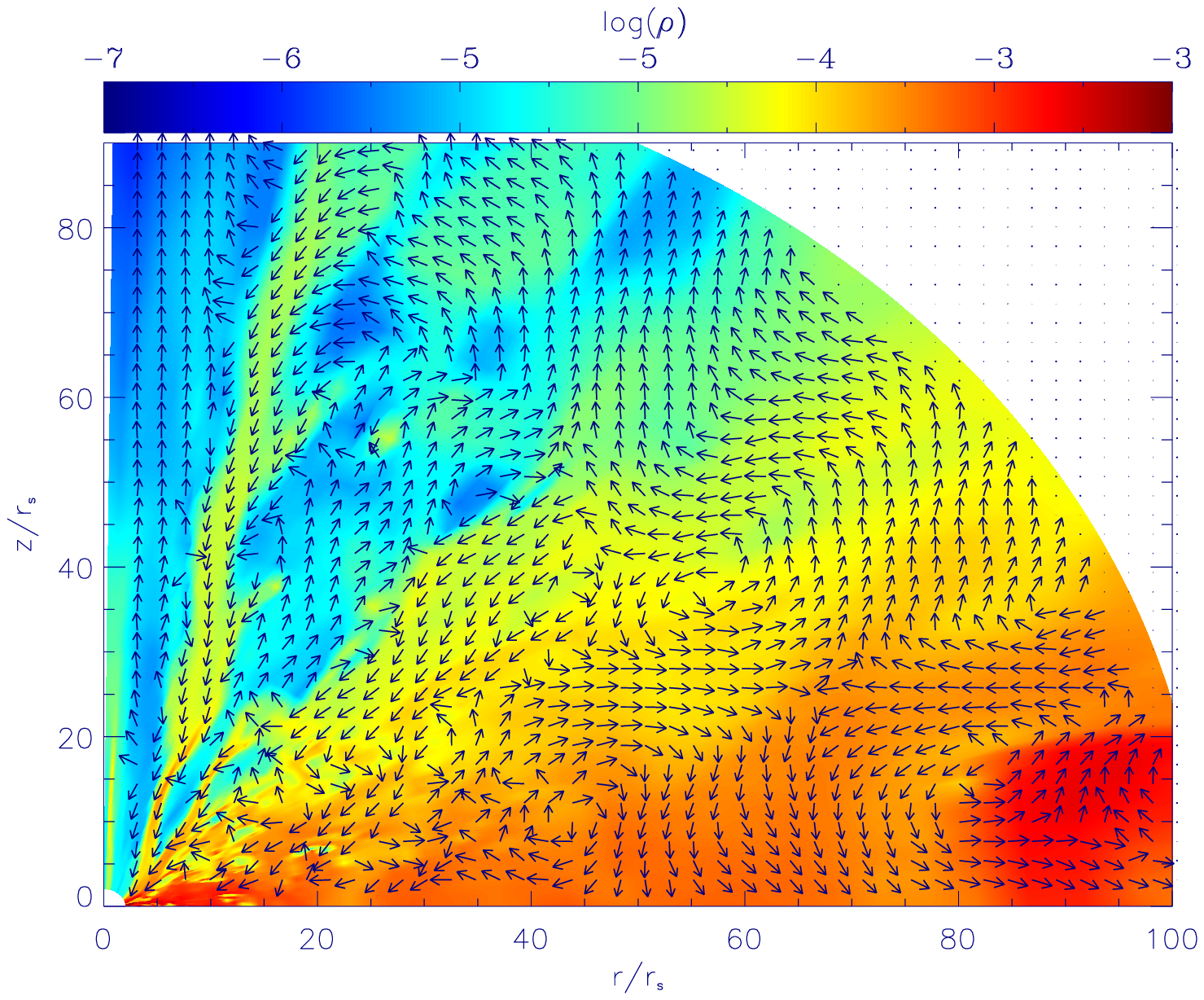}}}

\scalebox{0.78}[0.90]{\rotatebox{0}{\includegraphics[bb=61 358 494
716]{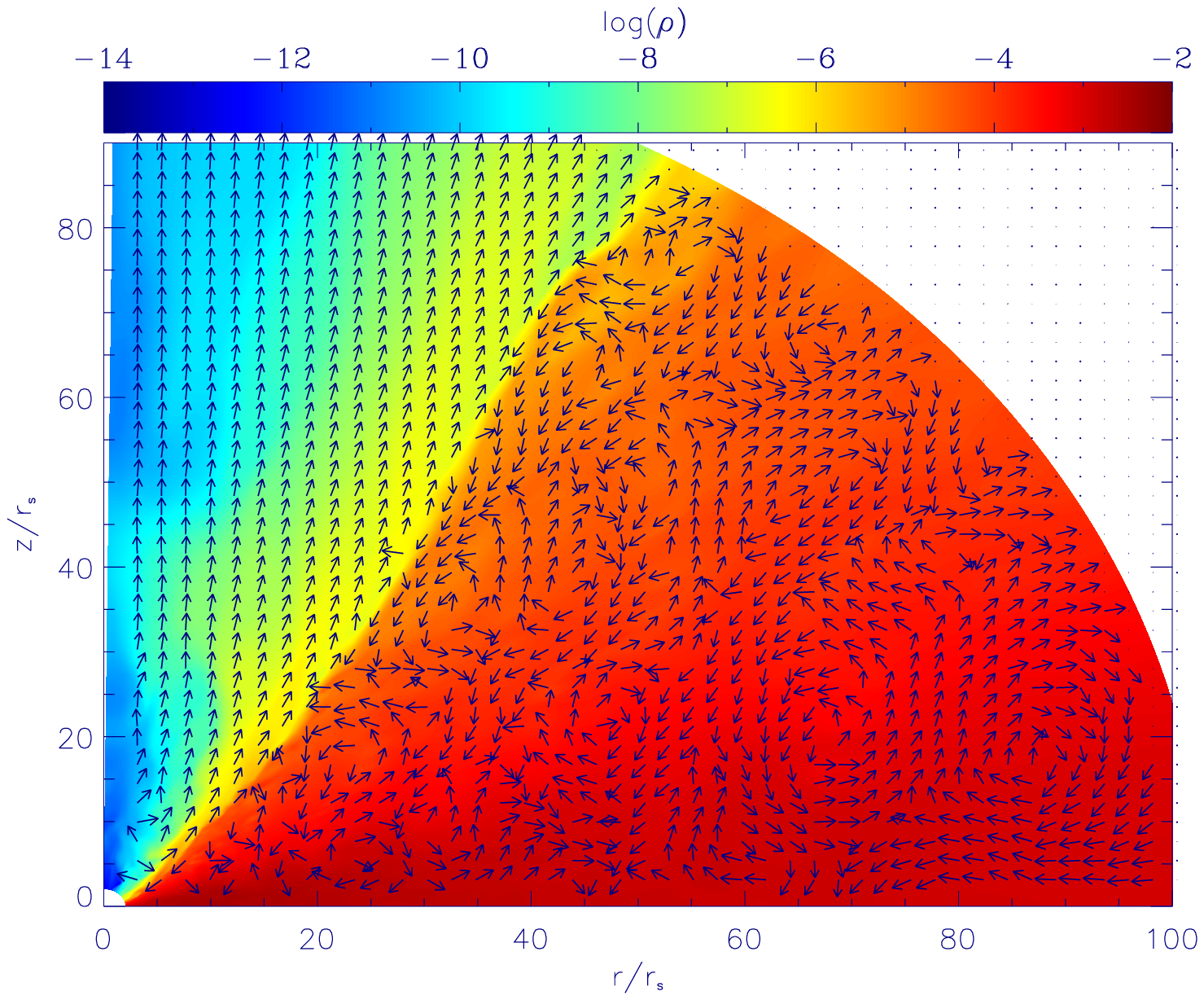}}}

\centering \caption{Snapshots of the logarithm of  density (colors),
overlaid with velocity vectors (arrows). Top: $t$ = 92.541 orbits of
Run 2b ($\alpha=0.1$); Bottom: 254.085 orbits of Run 5b
($\alpha=0.005$).} \label{fig 0}
\end{figure*}

The properties of all of the simulations are listed in Table 1,
where Columns 1 and 2 give the classification of our models and their
number; Column 3 gives the value of $\alpha$; Column 4 gives the mass injection rate
($\dot{m}_{\rm{input}}$) in units of the critical mass accretion
rate, $\dot{M}_{\rm{crit}}\equiv L_{\rm{Edd}}/c^2$; Columns 5, 6, and
7 give the computational domain in the $\theta$ direction, the components
of stress tensor, and the quantity of grid cells, respectively;
Column 8 gives the approximate dynamical time $t_s$ (all times in this
paper are reported in units of the orbital time at $r$=100$r_s$)
after which the accretion flows become quasi-steady; Column 9 gives
the final time $t_f$; and Columns 10 and 11 give the mass accretion rate onto the
black hole ($\dot{m}_{\rm{acc}}$) in the units of the critical mass
accretion rate and the luminosity ($L$) in the units of $L_{\rm
Edd}$, respectively. According to the component of stress tensor and
the angular range of computational domain, the simulations are
divided into two groups, namely Models A and B. Models A contains
only the $r\phi$ component of stress tensor, while Model B contains
both the $r\phi$  and $\theta\phi$ components.

Figure \ref{fig 0} displays a two-dimensional density distribution
overlaid with velocity vectors. The top and bottom panels are for
$t$=92.541 orbits of Run 2b and 254.085 orbits of Run 5b,
respectively.

\begin{figure*}

\scalebox{0.65}[0.65]{\rotatebox{0}{\includegraphics[bb=10 34 400
275]{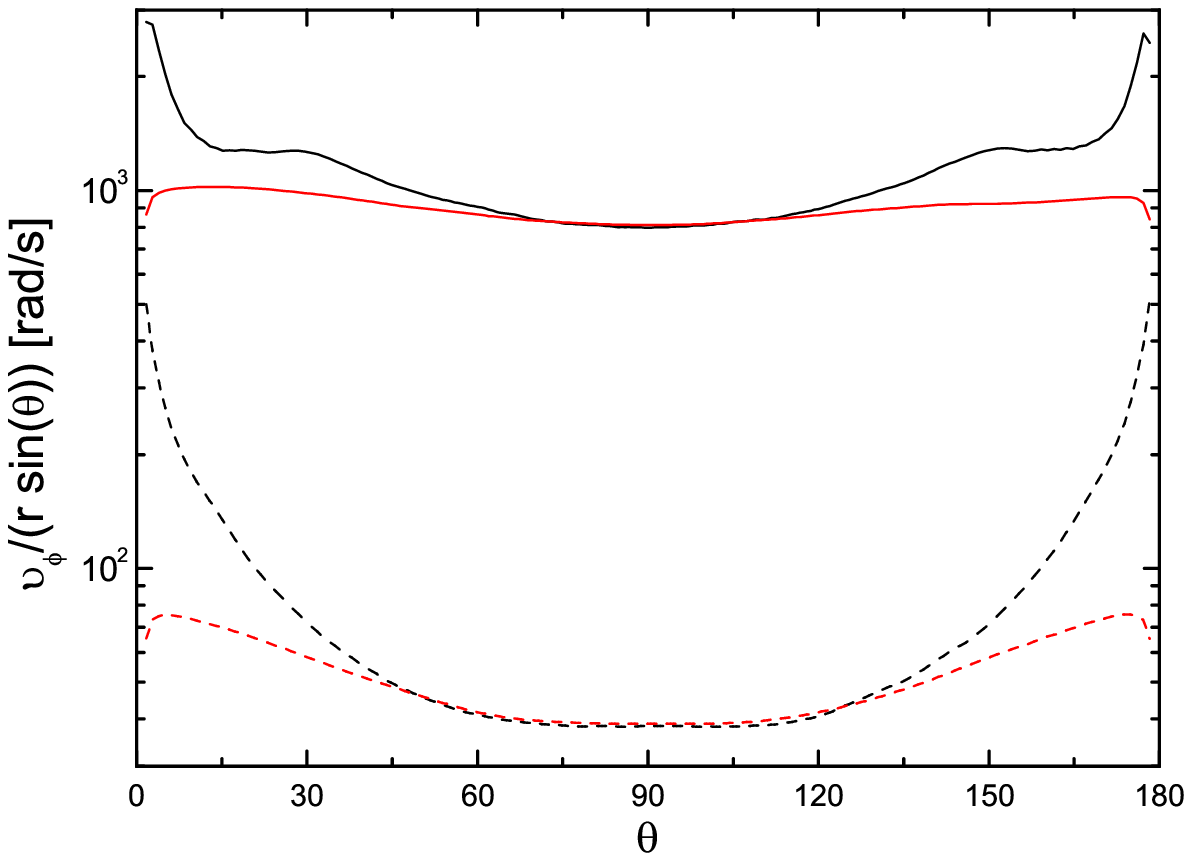}}}
\scalebox{0.65}[0.65]{\rotatebox{0}{\includegraphics[bb=10 34 450
275]{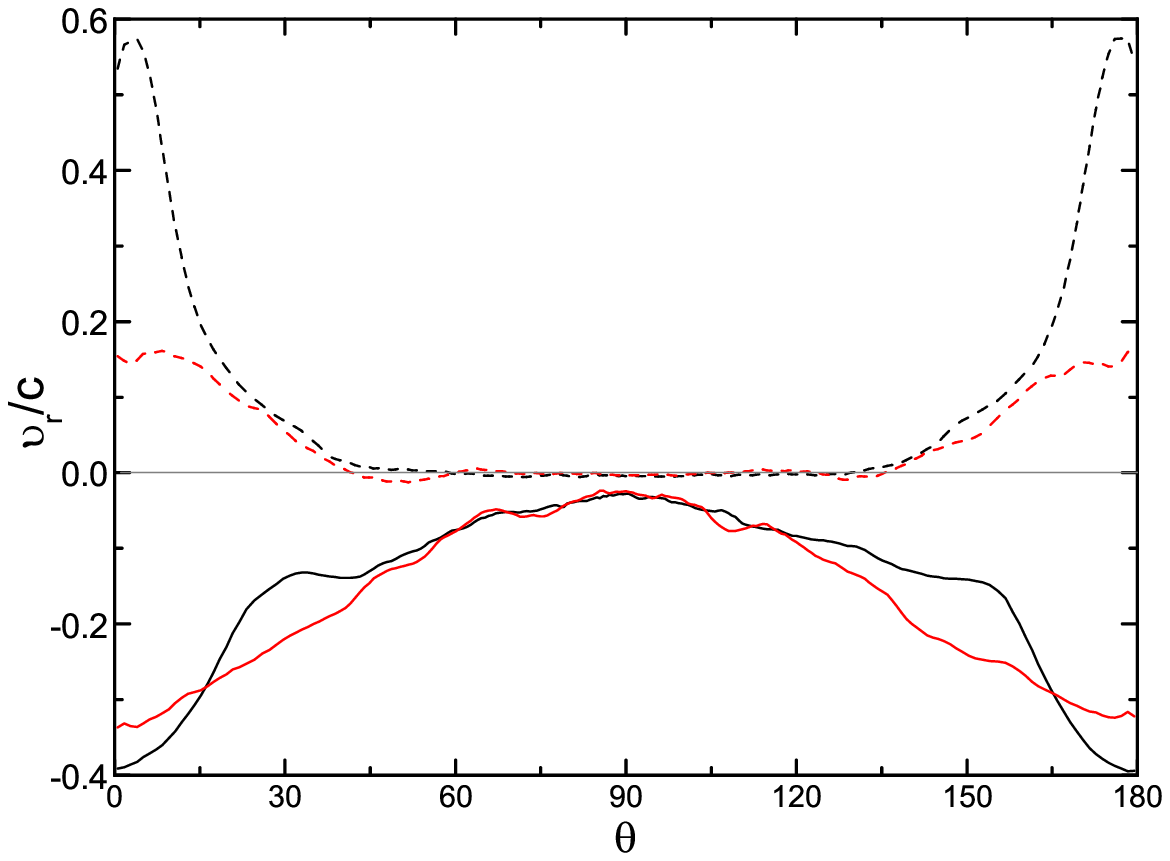}}}
\scalebox{0.65}[0.65]{\rotatebox{0}{\includegraphics[bb=10 34 400
290]{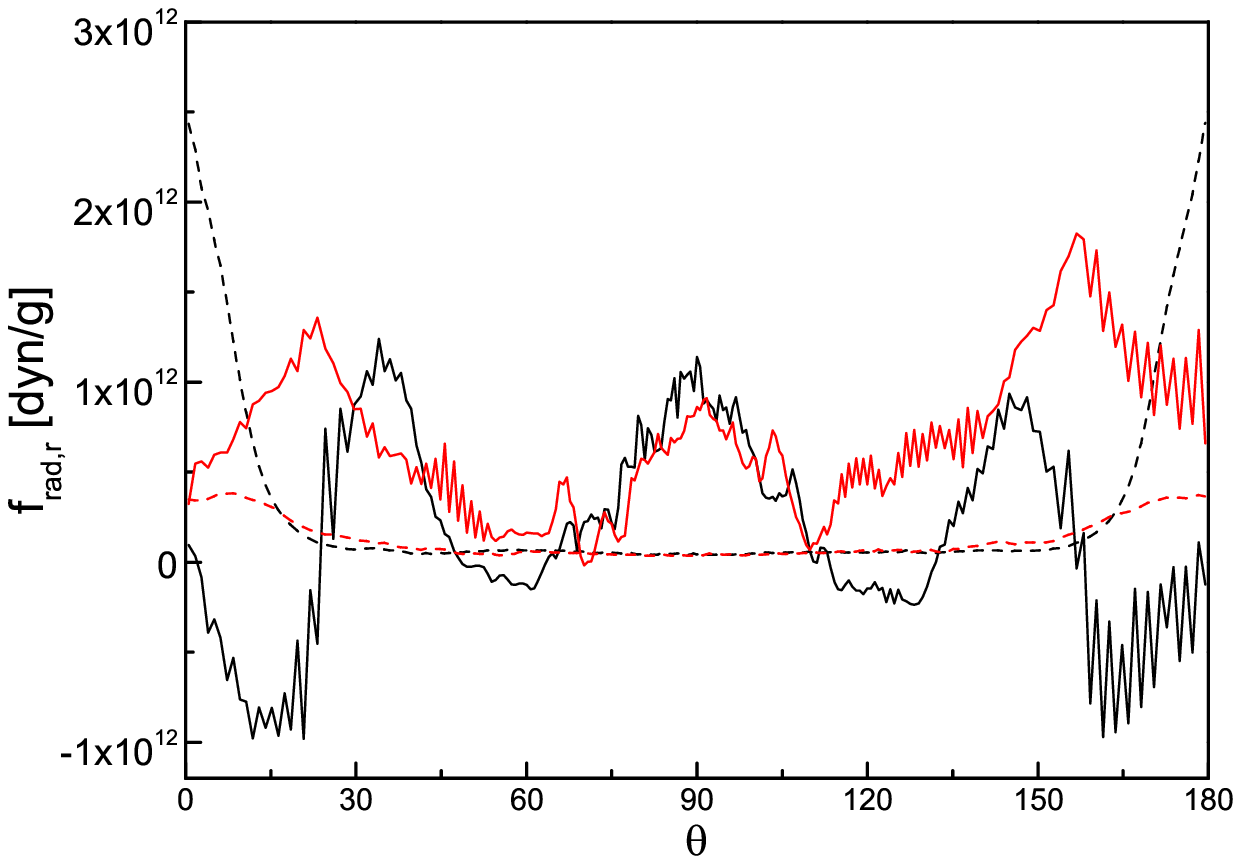}}}
\scalebox{0.65}[0.65]{\rotatebox{0}{\includegraphics[bb=10 34 400
290]{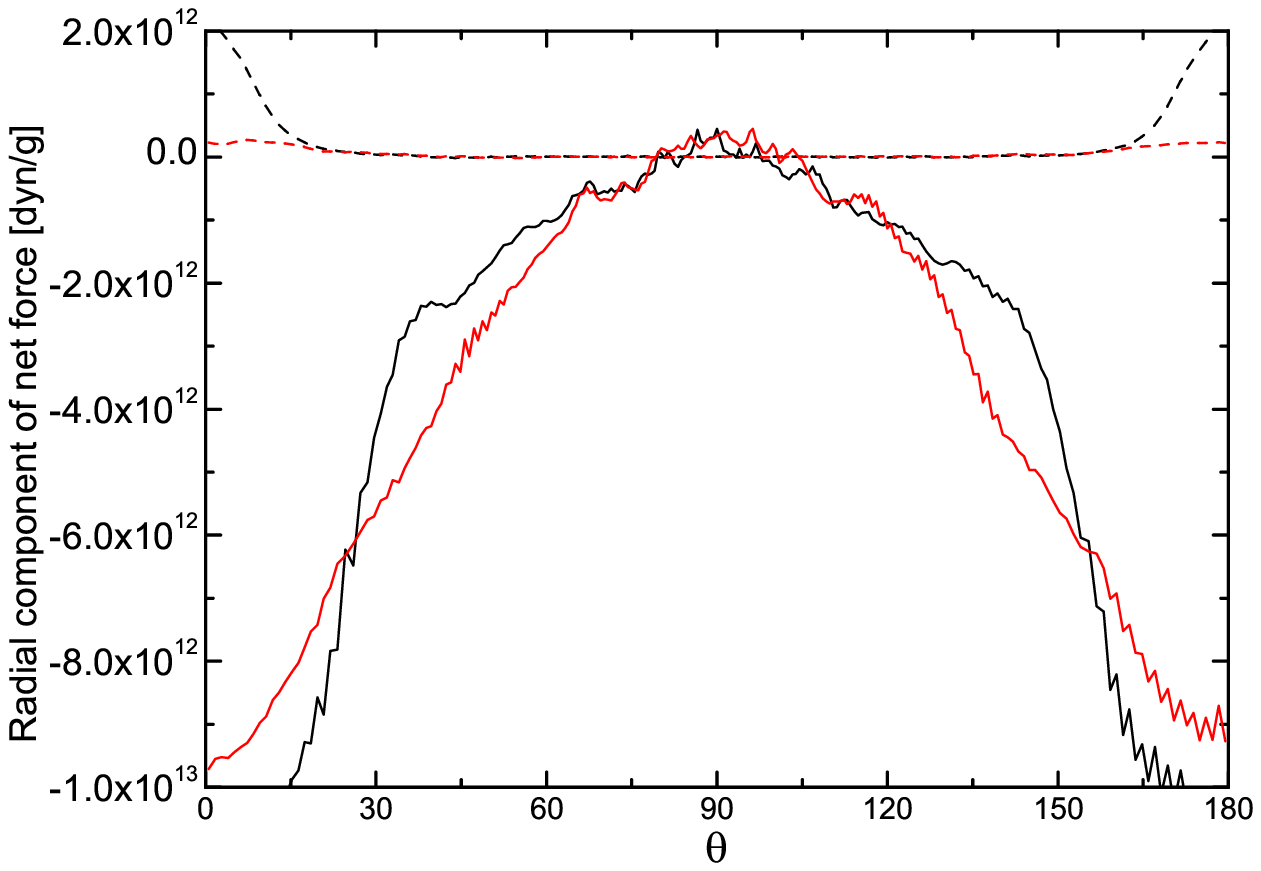}}}

\scalebox{0.65}[0.65]{\rotatebox{0}{\includegraphics[bb=10 34 400
290]{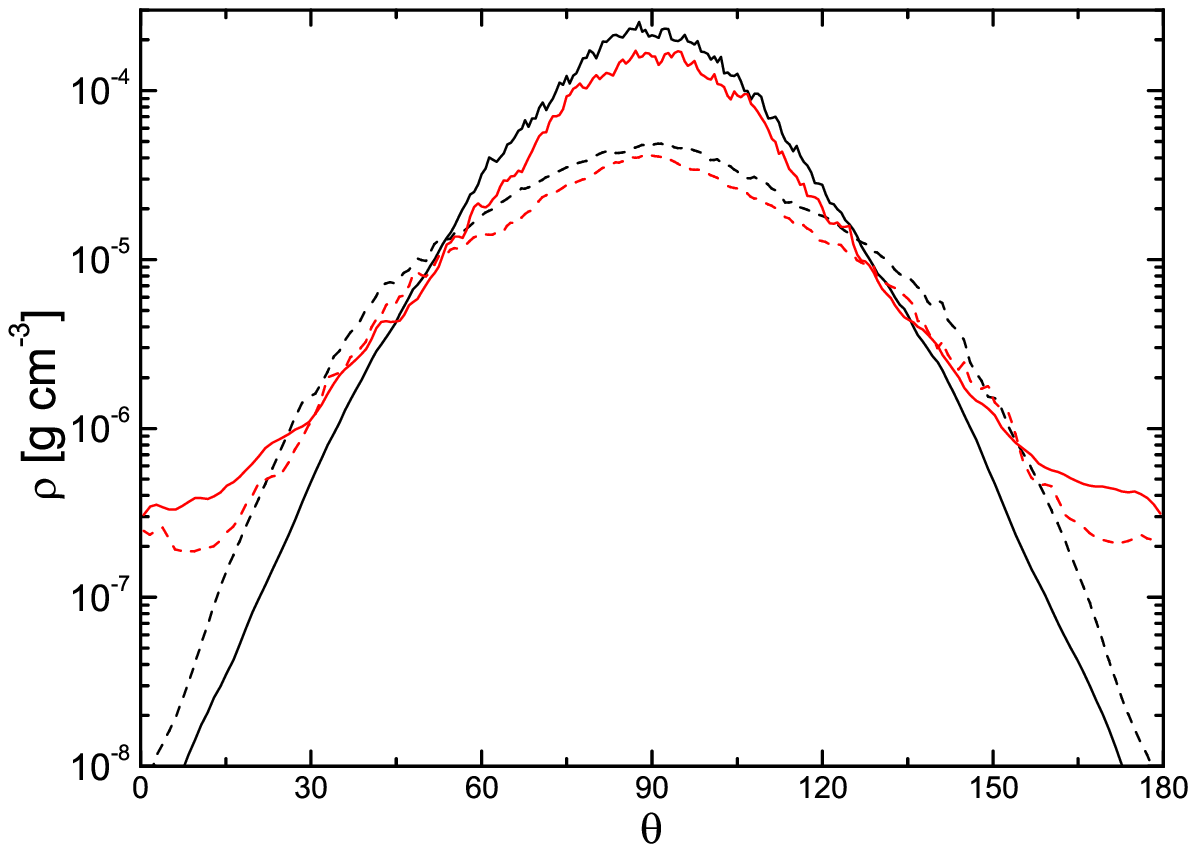}}}
\scalebox{0.65}[0.65]{\rotatebox{0}{\includegraphics[bb=10 34 450
290]{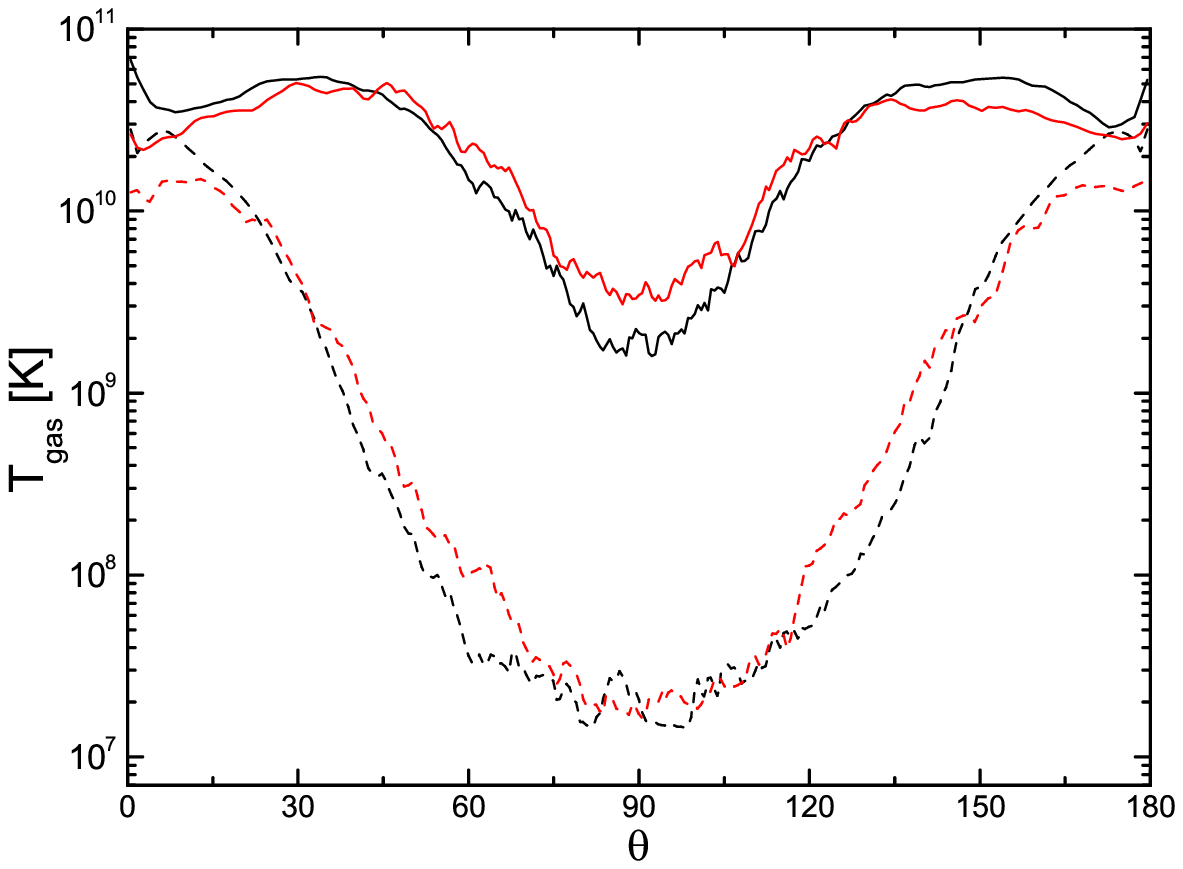}}}
\scalebox{0.65}[0.65]{\rotatebox{0}{\includegraphics[bb=10 34 400
290]{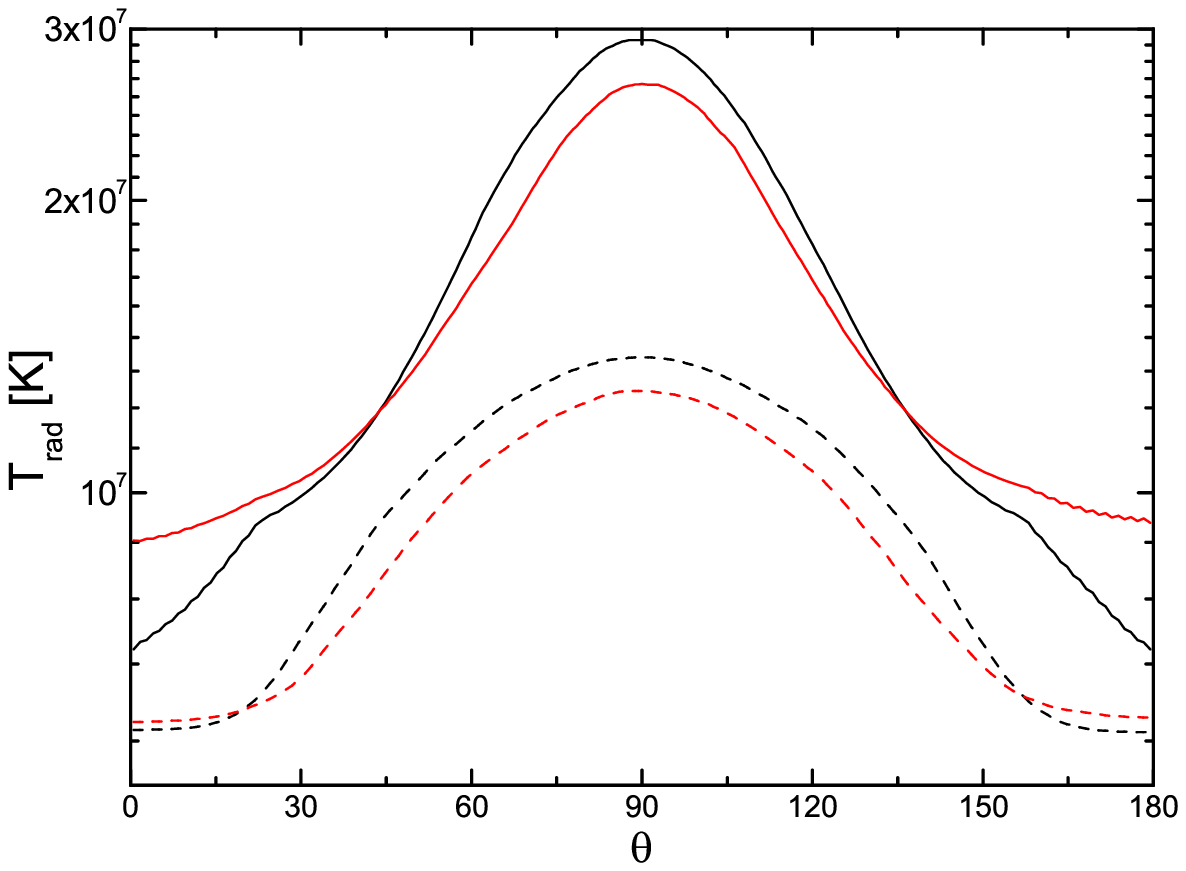}}}
\scalebox{0.65}[0.65]{\rotatebox{0}{\includegraphics[bb=10 34 400
290]{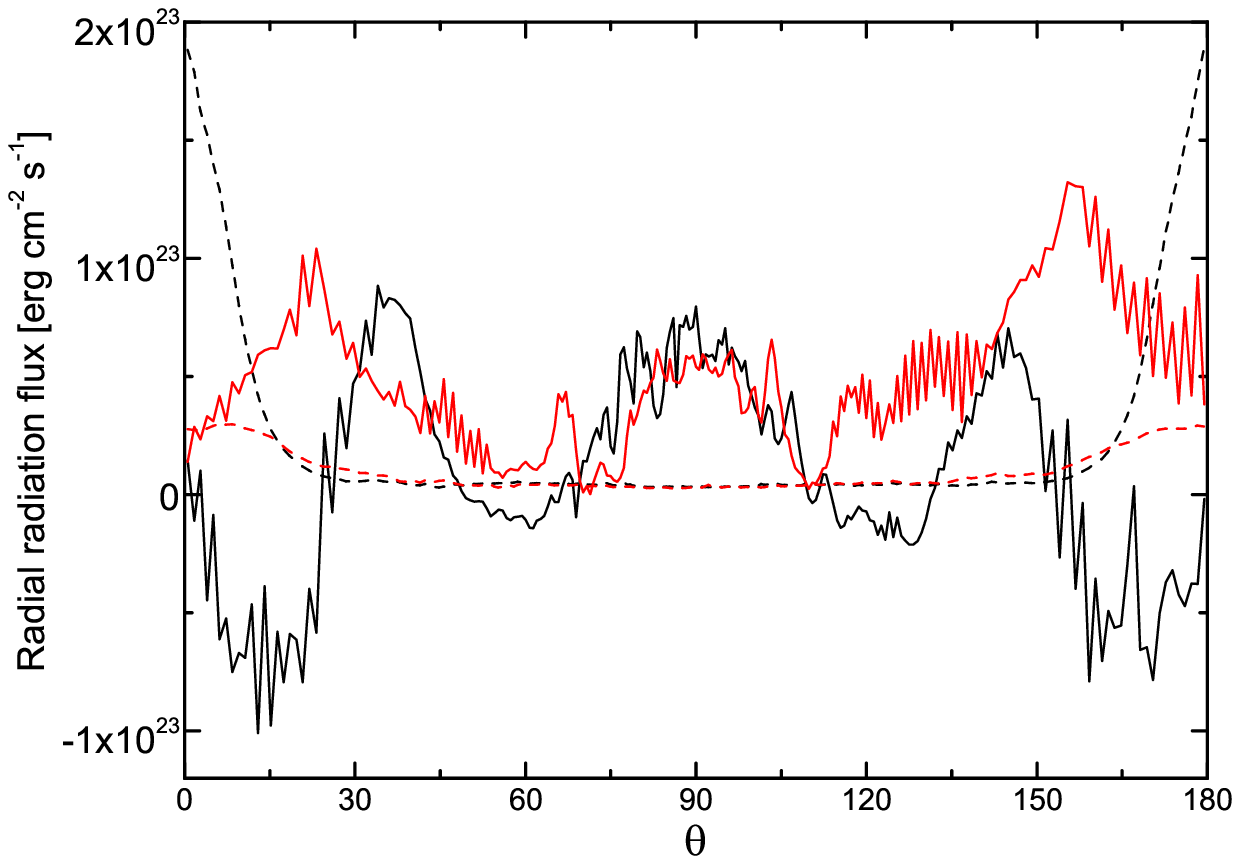}}}

\centering \caption{Angular profiles of a variety of time-averaged
variables for Run 1a ($T_{\theta\phi}=0$; black) and Run 1b (with
$T_{\theta\phi}\neq0$; red) at $r$=5$r_{s}$ (solid lines) and
30$r_{s}$ (dashed lines).} \label{fig 1}
\end{figure*}

\begin{figure}
\centering
\scalebox{0.65}[0.65]{\rotatebox{0}{\includegraphics[bb=40 28 390
300]{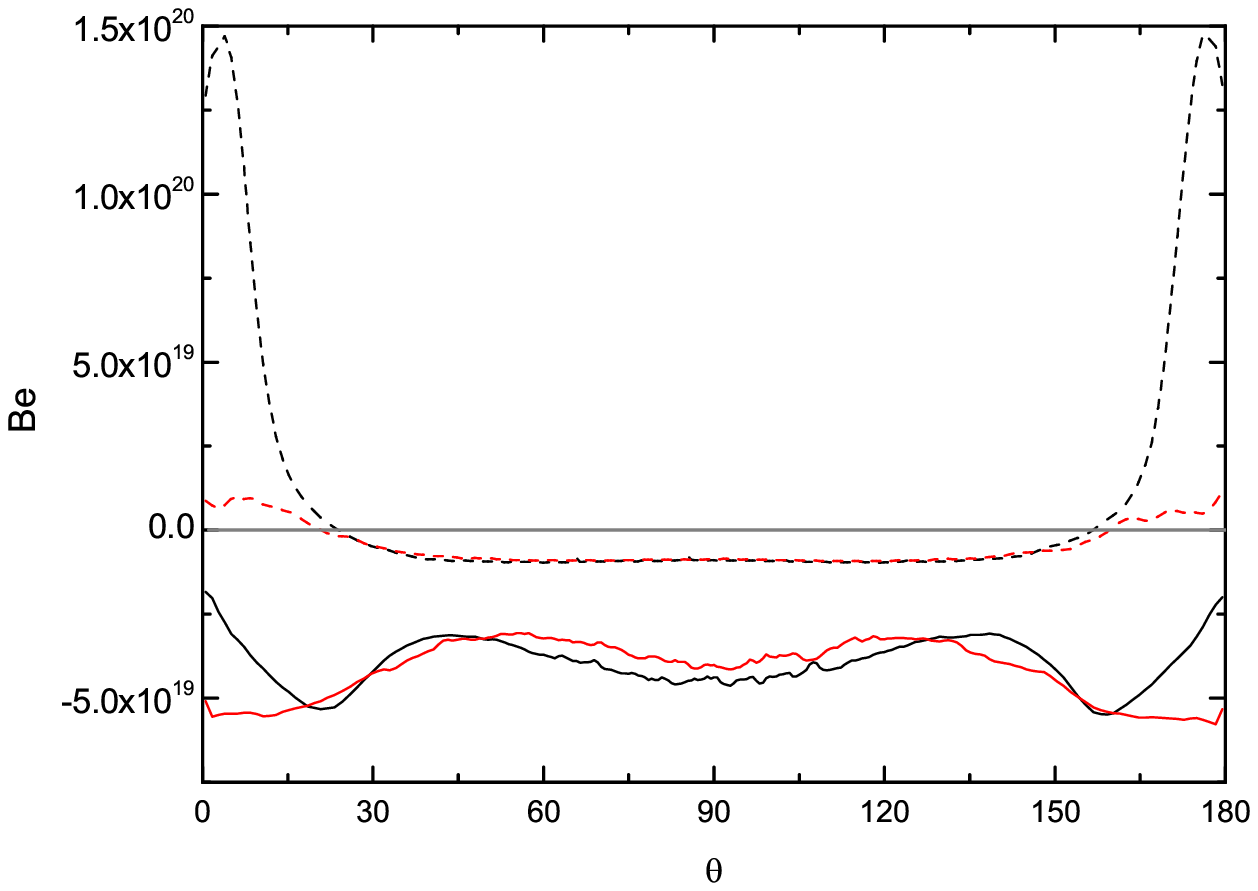}}}

\centering \caption{Angular profiles of time-averaged Bernoulli
parameters for Run 1a ($T_{\theta\phi}=0$; black) and Run 1b (with
$T_{\theta\phi}\neq0$; red) at $r$=5$r_{s}$ (solid lines) and
30$r_{s}$ (dashed lines). } \label{fig 2}
\end{figure}

\section{Results}

\subsection{The Effects of $T_{\theta\phi}$}

To investigate the effect of $T_{\theta\phi}$ on the dynamics of
accretion flow, we compare Models A and B. Figure \ref{fig 1} shows
the angular structure of time-averaged variables at $r=5 r_{s}$ and
30 $r_{s}$ for Run 1a and Run 1b with $\dot{m}_{\rm{input}}=1000$. In
this paper, all the time-averaged quantities are obtained by
averaging 100 data files within 23 orbits after the accretion flows
have achieved the quasi-steady state. In Figure  \ref{fig 1}, the black
solid ($r=5r_{s}$) and dashed ($r=30r_{s}$) lines correspond to Run
1a while the red solid ($r=5r_{s}$) and dashed ($r=30r_{s}$) lines
correspond to Run 1b. The figure shows obvious differences of the flow
structure, especially at the high-latitude region.

The angular profiles of angular velocity ($\upsilon_{\phi}/(r
\sin(\theta))$) show that the flow in Run 1a rotates faster than
that of Run 1b at the high-latitude region. Close to the equator, the
rotation of Run 1a is slightly slower than Run 1b. Because
$T_{\theta\phi}$ transports angular momentum between different
latitudes, for Run 1a the angular momentum between different
latitudes can not be transported, although the high-latitude flow
rotates faster than the low-latitude one. Therefore, it is seen that
the angular velocity increases from the equator to almost the axis
(although the angular velocity at the rotating axis is set to be
zero). When $\bm{T}_{\theta\phi}$ is included in Run 1b, the angular
momentum can be transported from the quickly rotating flows at high
latitude to the slowly rotating flows at low latitude. This is why
the discrepancy of the rotation velocity between the equator and the
high-latitude region is smaller in Run 1b than in Run 1a. The
discrepancy of the rotation velocity at the equator between Run 1a
and Run 1b is much smaller than in the high-latitude region, since the
density at the high-latitude region is much smaller than in the equator.

The angular profiles of radial velocity ($v_{r}$) show obvious
differences between Run 1a and Run 1b, especially at the
high-latitude region. The positive value of $v_{r}$ indicates that
the flows are outflows, while the negative value indicates the
inflow nature of flows. At large radii such as $\sim 30r_s$, the
high-latitude flow (the flow within $30^\circ$ from the axis) is
mainly outflow. The speed of the high-latitude outflow of Run 1a is
higher than that of Run 1b. At small radii such as $r<5r_s$, the
flow is inflow at all latitudes for both models. The angular
profiles of radial velocity agree with the angular profiles of
radial forces, as shown in Figure \ref{fig 1}. The radial radiation
force dominates the radial component of net force at the
high-latitude region at $\sim 30 r_s$; therefore, it is the dominant
force changing the angular distribution of the radial velocity
there. At small radii, gravity is the dominant force.

The angular profiles of density ($\rho$) of Run 1a and Run 2b are
similar, with the maximum density located at the equator. Compared
with Run 1a, the density of the high-latitude region of Run 1b is
higher, while at the equator is slightly lower. The angular
profiles of gas temperature ($T_{\rm gas}$) show that the disk of
Run 1a and Run 1b has nearly the same gas temperature at $30r_s$. For
Run 1b the gas temperature of high-latitude outflow is lower than
that of Run 1a. For the high-latitude outflow, the lower the gas
temperature, the higher the density.

\begin{figure*}
\scalebox{0.67}[0.67]{\rotatebox{0}{\includegraphics[bb=31 28 420
279]{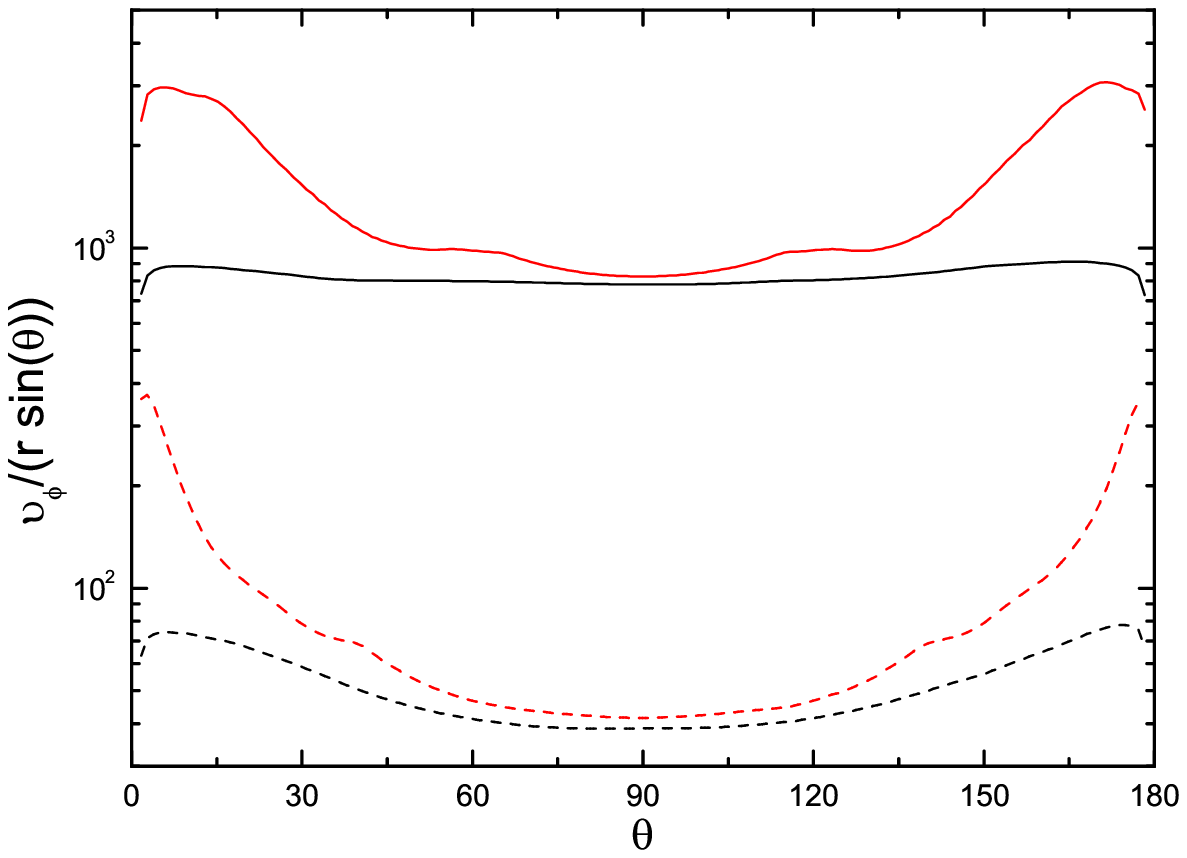}}}
\scalebox{0.67}[0.67]{\rotatebox{0}{\includegraphics[bb=35 28 450
270]{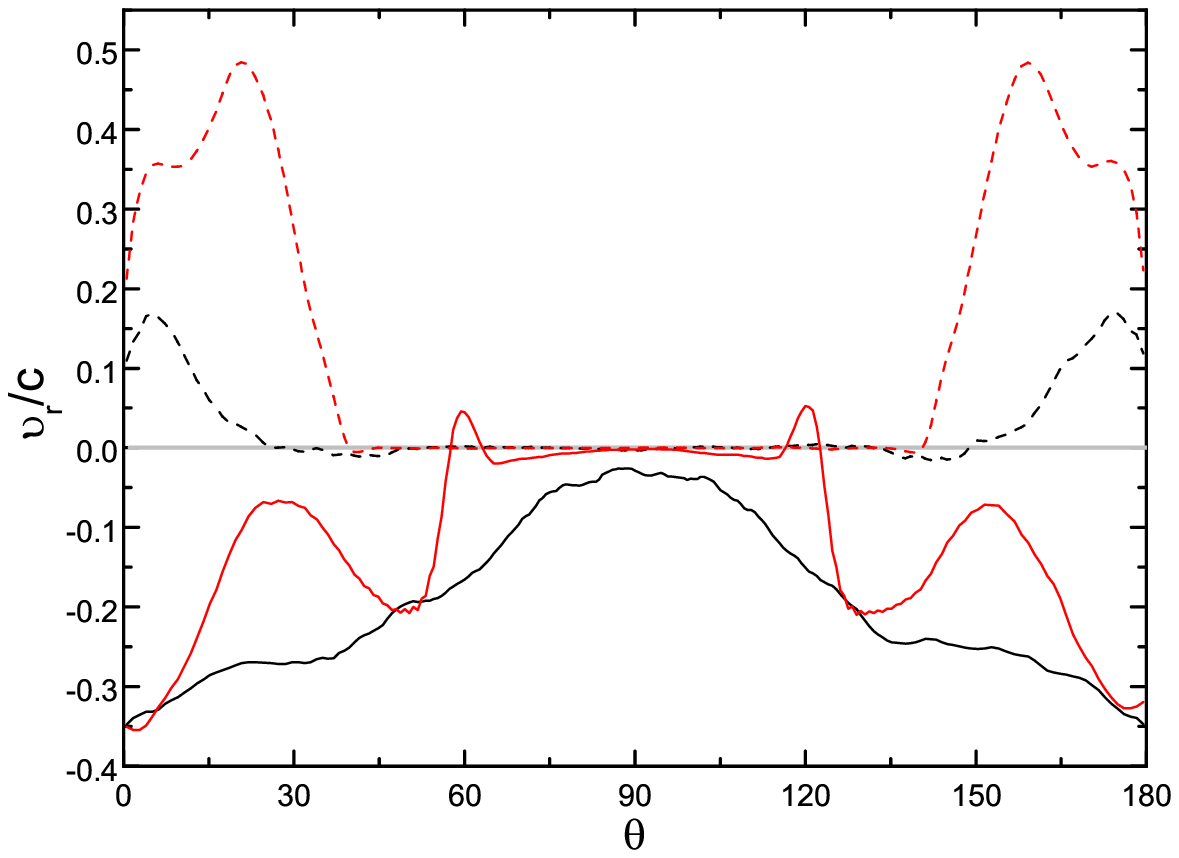}}}
\scalebox{0.67}[0.67]{\rotatebox{0}{\includegraphics[bb=31 28 400
270]{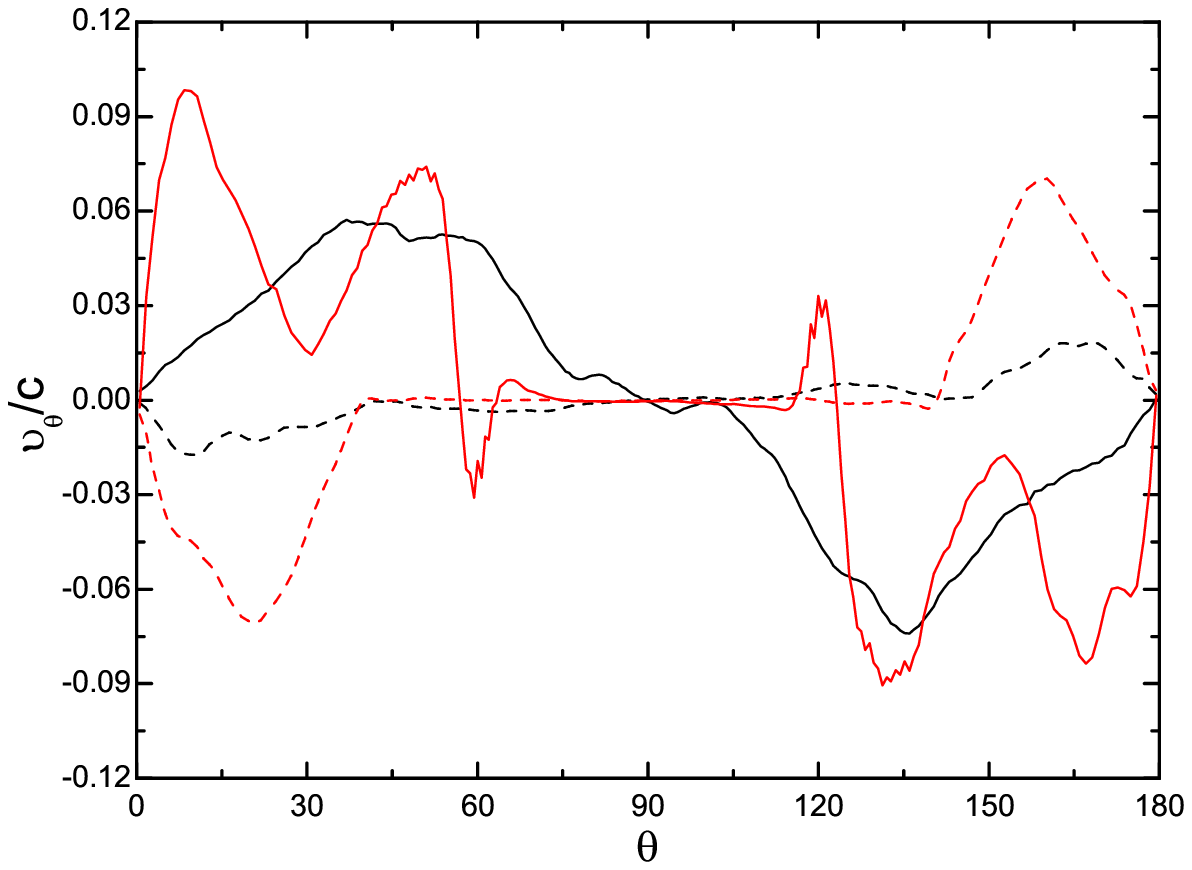}}}
\scalebox{0.67}[0.67]{\rotatebox{0}{\includegraphics[bb=15 28 400
284]{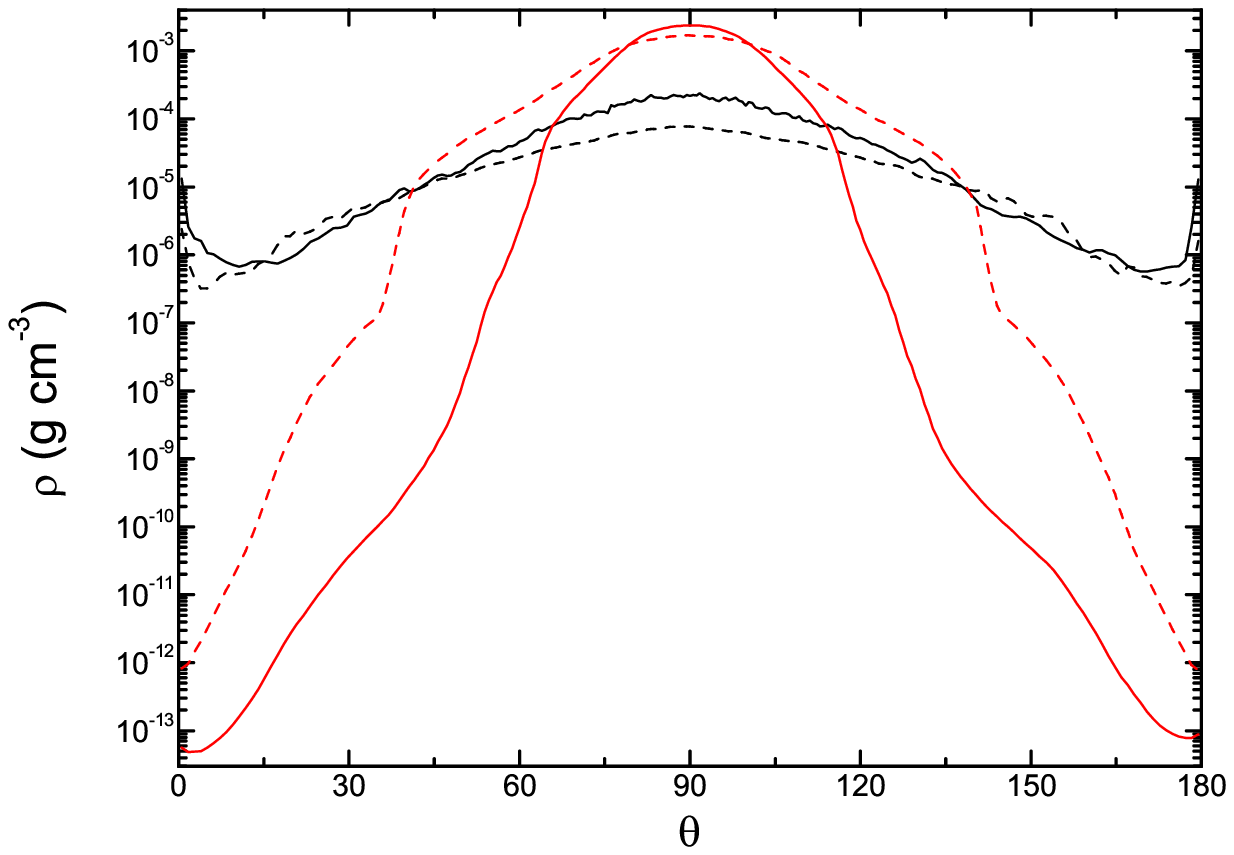}}}

\scalebox{0.67}[0.67]{\rotatebox{0}{\includegraphics[bb=31 28 420
284]{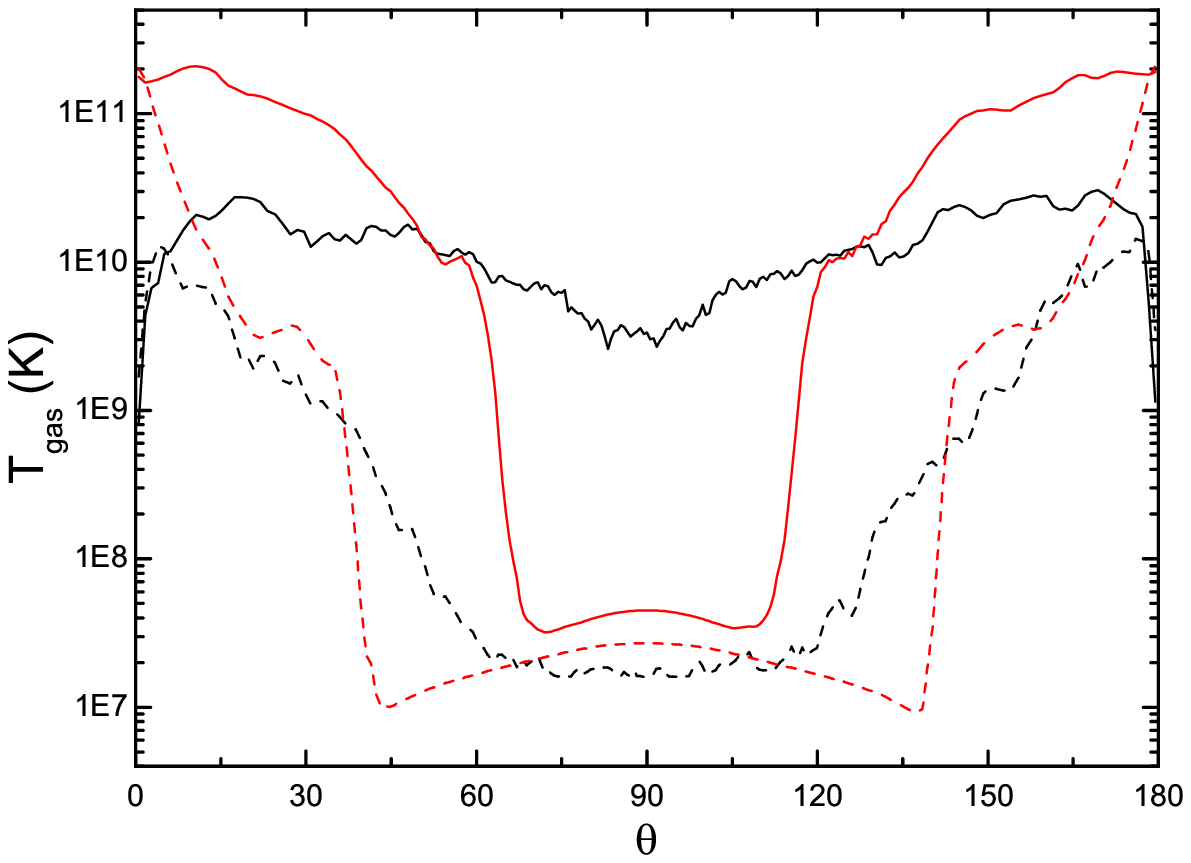}}}
\scalebox{0.67}[0.67]{\rotatebox{0}{\includegraphics[bb=35 28 450
284]{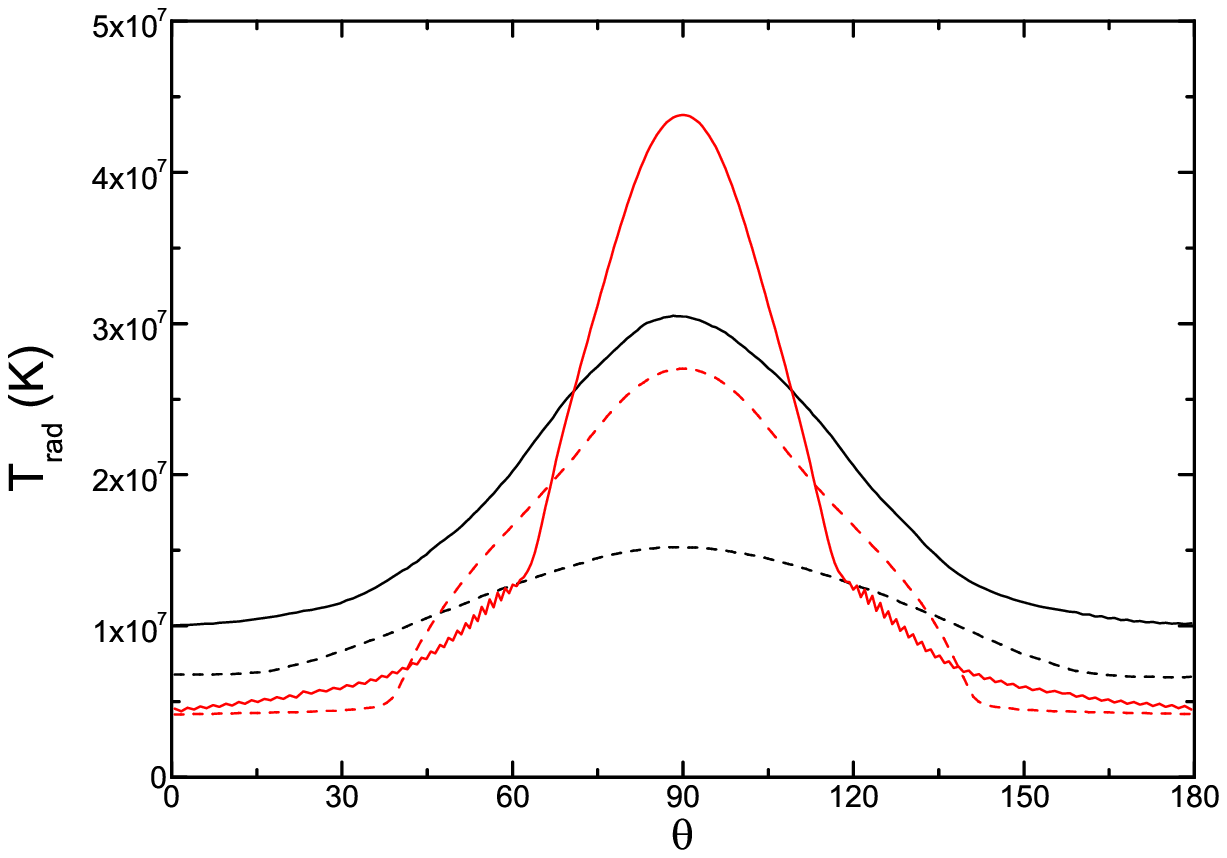}}}

\centering \caption{Angular profiles of a variety of time-averaged
variables from Run 2b ($\alpha=0.1$; black) and Run 5b ($\alpha=0.005$;
red) at $r=5r_s$ (solid lines) and $30r_s$ (dashed lines).}
\label{fig 3}
\end{figure*}

\begin{figure}
\scalebox{0.66}[0.66]{\rotatebox{0}{\includegraphics[bb=25 28 400
300]{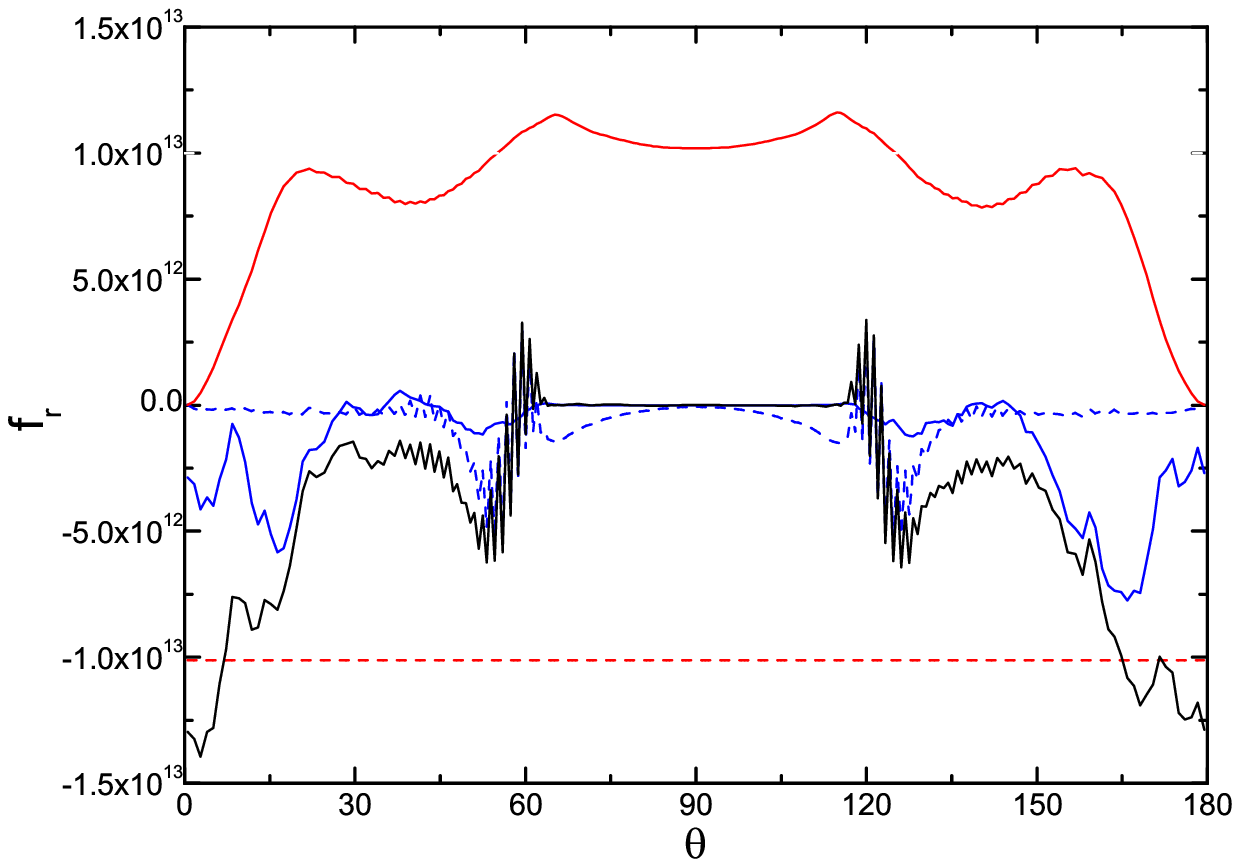}}}
\scalebox{0.66}[0.66]{\rotatebox{0}{\includegraphics[bb=25 28 400
300]{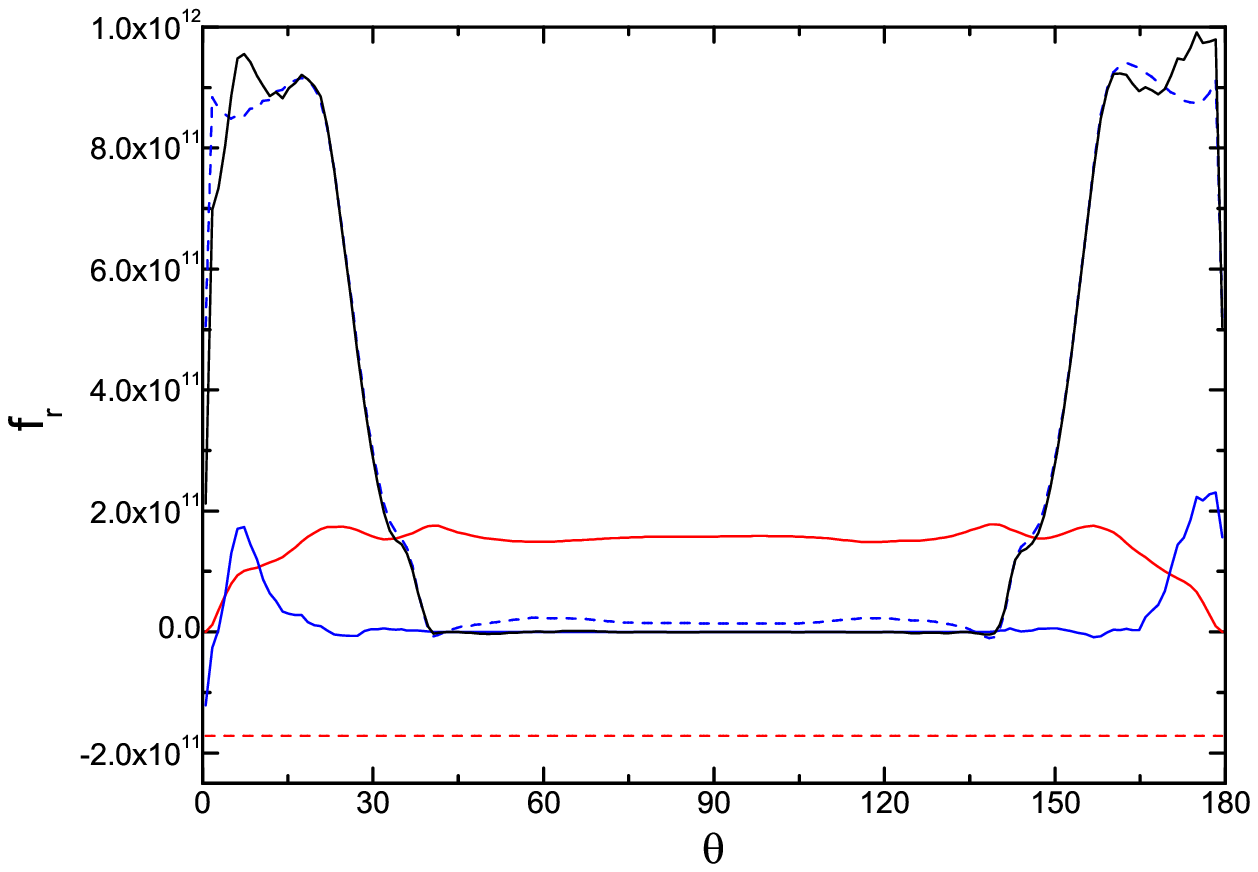}}}

\centering \caption{Angular distribution of the radial forces per
unit mass for Run 5b at $r=5r_s$ (top panel) and $30r_s$ (bottom
panel). The radial forces include gravity (red dashed line),
centrifugal force (red solid line), radiation force (blue dashed
line), gas-pressure gradient force (blue solid line), and their sum
(black solid line). } \label{fig 4}
\end{figure}

\begin{figure}
\scalebox{0.66}[0.66]{\rotatebox{0}{\includegraphics[bb=25 28 400
300]{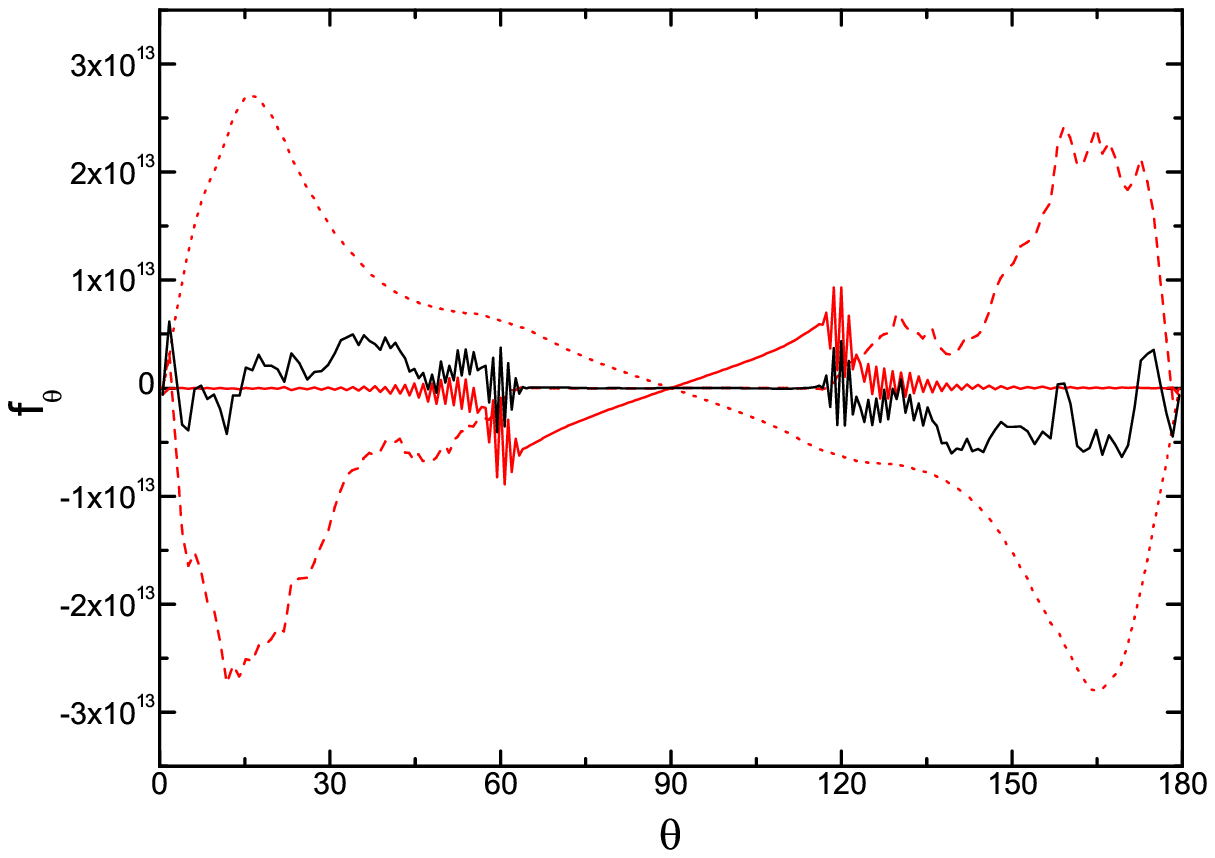}}}
\scalebox{0.66}[0.66]{\rotatebox{0}{\includegraphics[bb=25 28 400
300]{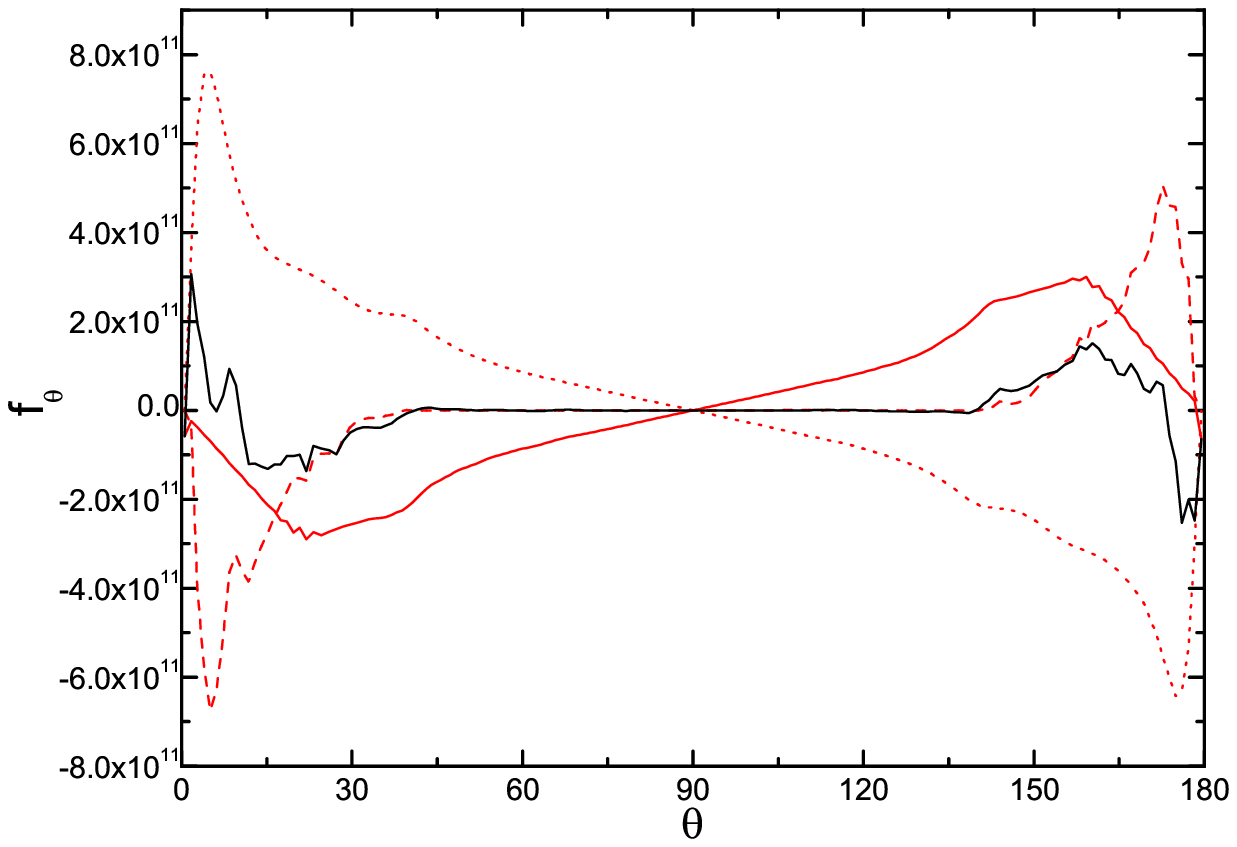}}}

\centering \caption{Angular distribution of the angular forces per
unit mass for Run 5b at $r=5r_s$ (top panel) and $30r_s$ (bottom
panel). The angular forces include the angular component of
radiation force (red solid line), centrifugal force (red dotted
line), gas-pressure gradient force (red dashed line), their sum (is
shown by the black solid line.) } \label{fig 5}
\end{figure}

The angular profiles of the ``radiation" temperature ($T_{\rm rad}
\equiv (E_0/a)^{\frac{1}{4}}$, where $a$ is the radiation constant)
show that compared with Run 1a, the radiation temperature for Run 1b
is lower near the equator and higher near the rotating axis . For
the super-critical accretion flow, the radiation force (including
also the gradient of the radiation pressure) is the dominant force
driving the high-latitude outflow. The radiation force acting on
unit mass is given by $\bm{f}=\frac{\chi}{c\rho}\bm{F}_{0}$, where
the flux-mean opacity $\chi$ is the sum of components due to
absorption and scattering. The scattering in the high-latitude
outflow is the dominant factor for the opacity so that $\chi \propto
\rho$ and $\bm{f}\propto \bm{F}_{0}$. Here we employ the
approximation and have $\bm{F}_{0}=-\frac{c\lambda}{\chi} \nabla
E_{0}$. In the optically thin limit, $|\bm{F}_{0}| = cE_{0}$. In
general, the high-latitude outflow has high gas temperature and low
density, so the flow is optically thin. For Run 1a, the opacity of
outflow near the rotating axis is close to the optically thin limit,
while for Run 1b the opacity of outflow is away from the optically
thin limit because of the higher density. Therefore, although Run 1a
has lower radiation energy density at the high-latitude region than
Run 1b, the radiation flux of Run 1a is larger than that of Run 1b,
which is shown by the angular profiles of radial radiation flux in
Figure \ref{fig 1}. Therefore, at the high-latitude region the
radiative force of Run 1a is larger, exceeding that of Run 1b.
Compared with Run 1a,  the density of high-latitude outflow for Run
1b is higher so that the driving force of high-latitude outflow is
weaker.

For the strict steady state and for inviscid hydrodynamic flow, the
Bernoulli parameter Be (Be $\equiv \upsilon^{2}/2 +\gamma
p/(\gamma-1)\rho-GM/(r-r_{s})$) is conserved along the streamline.
Therefore, the positive sign of Be is often used to be the necessary
condition for the outflow to escape to infinity. However, in our
case, these conditions are not satisfied, thus Be is no longer
conserved (e.g., Yuan et al. 2012). In fact, the initial result from
our ongoing work indicates that Be can increase along the trajectory
of outflow elements (F.Yuan et al. 2013, in preparation). This means
that an outflow with a negative Be can also potentially escape to
infinity. Despite these uncertainties, we still show in Figure
\ref{fig 2} the angular distribution of the time-averaged Bernoulli
parameter Be because its value may still play some role in
determining the properties of outflow. We can see that at the region
close to the axis ($\theta < 25^{\circ}$) and $r=30r_s$, the value
of Be in Run 1b is about one order of magnitude lower than that in
Run 1a. From Figure \ref{fig 1}, we can see that including
$T_{\theta\phi}$ for the polar outflows, rotational speed, radial
speed and temperature decrease can decrease the specific Be of
outflow. In general, the value of Be is larger close to the axis.
When compared with the hydro and MHD numerical simulations of hot
accretion flow presented in Yuan et al. (2012), it is interesting to
note that the angular distribution of Be is more similar to the MHD
case (their Model D in Yuan et al. (2012)) rather than to the hydro
case (their Models A, B, and C). This is perhaps because the
radiation acceleration in the $r$ direction is much stronger in the
region close to the axis than in other directions. In the case of
Models A, B, C in Yuan et al. (2012), there is no such  force. But
in the MHD case, the magnetic pressure force plays a similar role
with the radiation force here.

\subsection{The Effect of the Viscous Parameter $\alpha$}

Figure \ref{fig 3} shows the angular distribution of the
time-averaged flow for Model B (with $\dot{m}_{\rm{input}}=3000$;
i.e., Run 2b and 5b) at $r=5 r_s$ and $r=30 r_s$, respectively. We
can see that the viscosity parameter $\alpha$ obviously affects the
angular distribution of accretion flow. The angular profiles of
angular velocity ($v_{\phi}/(r \text{sin}(\theta$))) show a large
difference close to the axis because of the different $\alpha$. When
$\alpha$ is smaller, the flow rotates faster. A smaller $\alpha$
results in a smaller $\bm{T}_{\theta\phi}$; thus, the angular
momentum cannot be efficiently transported from high to low
latitudes. The angular profile of the radial velocity ($v_{r}/c$) is
different for different $\alpha$. The speed of the high-latitude
outflow in Run 15b is higher than that of Run 2b. At $r=30 r_s$, the
high-speed outflow is constrained to be within $30^{\circ}$ from the
axis. With the decrease of $\alpha$, the range of $\theta$ within
which high-latitude outflow takes place becomes larger toward the
equator. At $r=5 r_s$, the radial velocity is negative over most
$\theta$.

The plots of angular profiles of density, gas temperature, and
``radiation'' temperature show that these three quantities change
rapidly with $\theta$ for Run 5b. With the decrease of $\alpha$, the
density and radiation energy density concentrate toward the equator,
and the flow can be divided into two areas with different
temperature, i.e.,a low-temperature area near the equator and a high-temperature area near the axis.

\begin{figure*}
\scalebox{1.4}[1.4]{\rotatebox{0}{\includegraphics[bb=40 1 220
300]{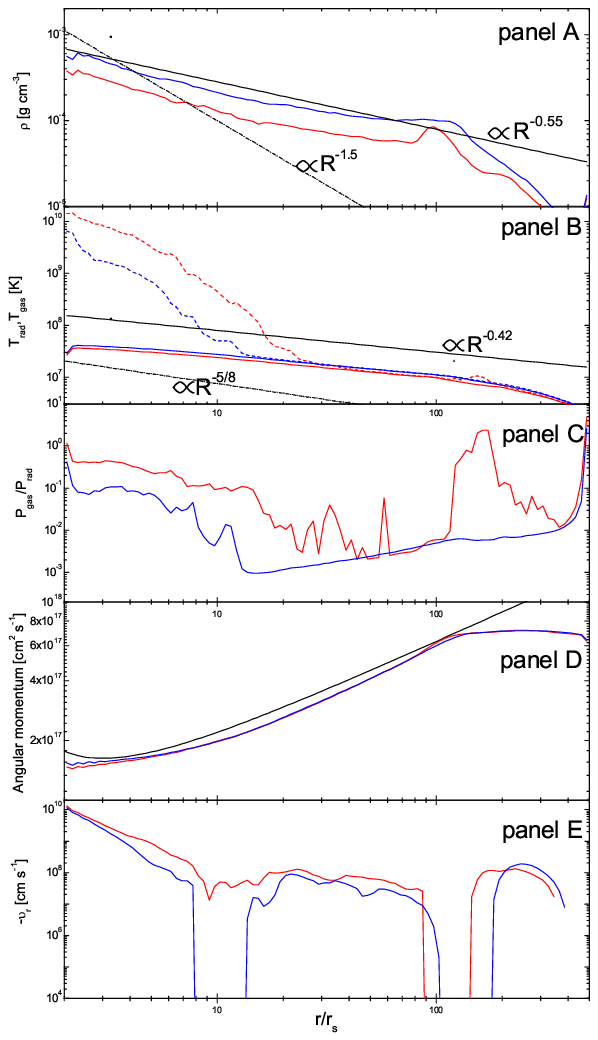}}}
\scalebox{1.4}[1.4]{\rotatebox{0}{\includegraphics[bb=40 1 220
300]{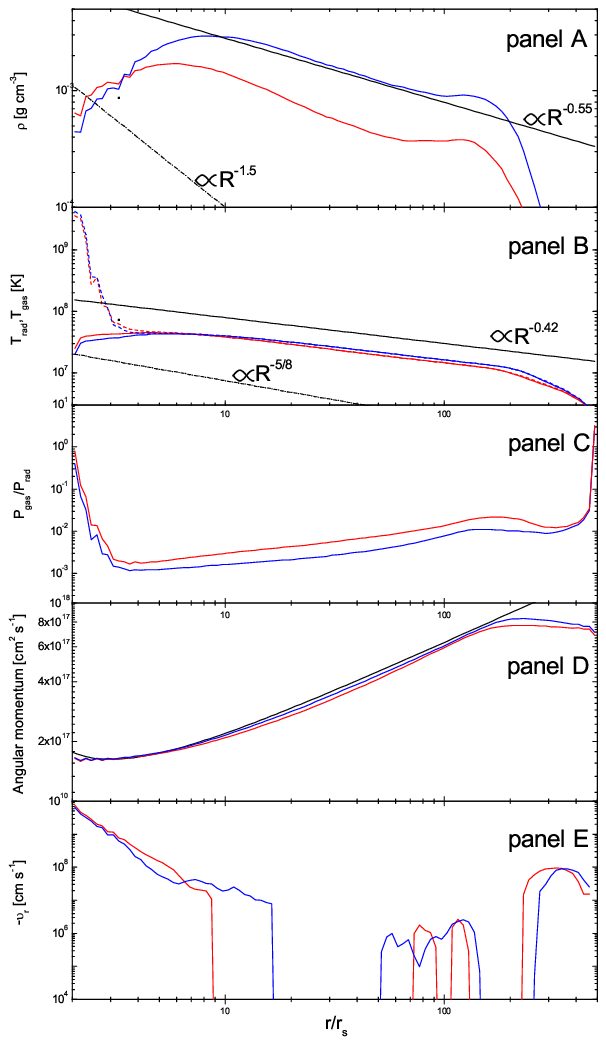}}}

\centering \caption{Radial structure of some time-average
quantities. The left panel is for Run 2b ($\alpha=0.1$; red line)
and Run 3b ($\alpha=0.05$; blue line), while the right panel is for
Run 4b ($\alpha=0.01$; red line) and Run 5b ($\alpha=0.005$; blue
line). The plots from top to bottom are for density, gas and
radiation temperature, the ratio of gas pressure to radiation
pressure, specific angular momentum, and radial velocity,
respectively. The dot-dashed lines denote the profile of the
one-dimensional solution. The black solid line is to guide our eyes.
In panel (D), the black solid line indicates the Keplerian angular
momentum.} \label{fig 6}
\end{figure*}

\begin{figure}

\scalebox{1.35}[1.4]{\rotatebox{0}{\includegraphics[bb=50 62 210
315]{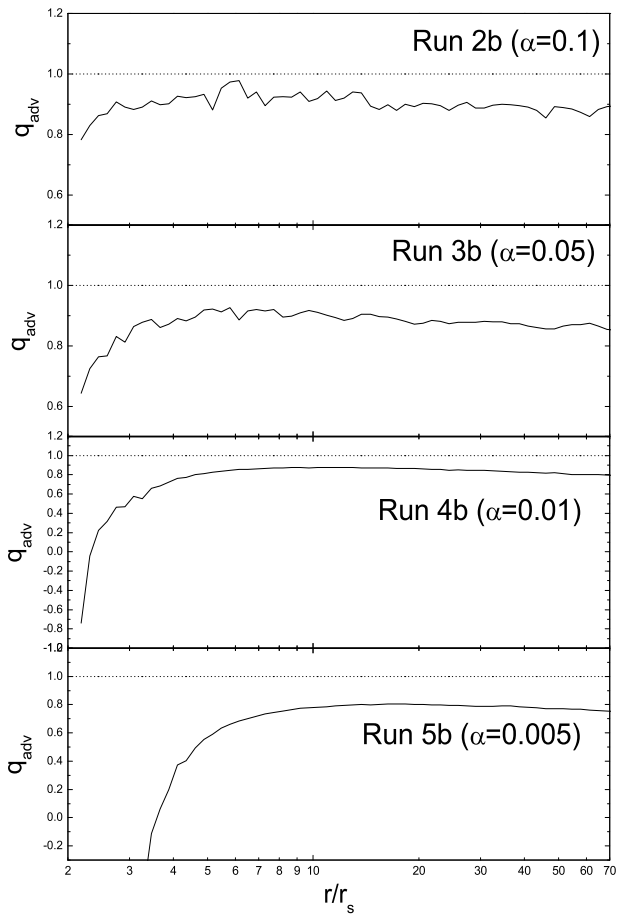}}} \vspace{0.5cm}

\centering \caption{Time-averaged advection factor $q_\text{adv}$
(cf. Equation (3)) near the equator for the models of
$\dot{m}_{\text{input}}=3000$. The solution is averaged over an
angle between $\theta=84^\text{o}$ and $\theta=96^\text{o}$. Dotted
lines indicate $q_\text{adv}$=1.} \label{fig 7}

\end{figure}

\begin{figure*}
\scalebox{1.3}[1.3]{\rotatebox{0}{\includegraphics[bb=30 25 418
318]{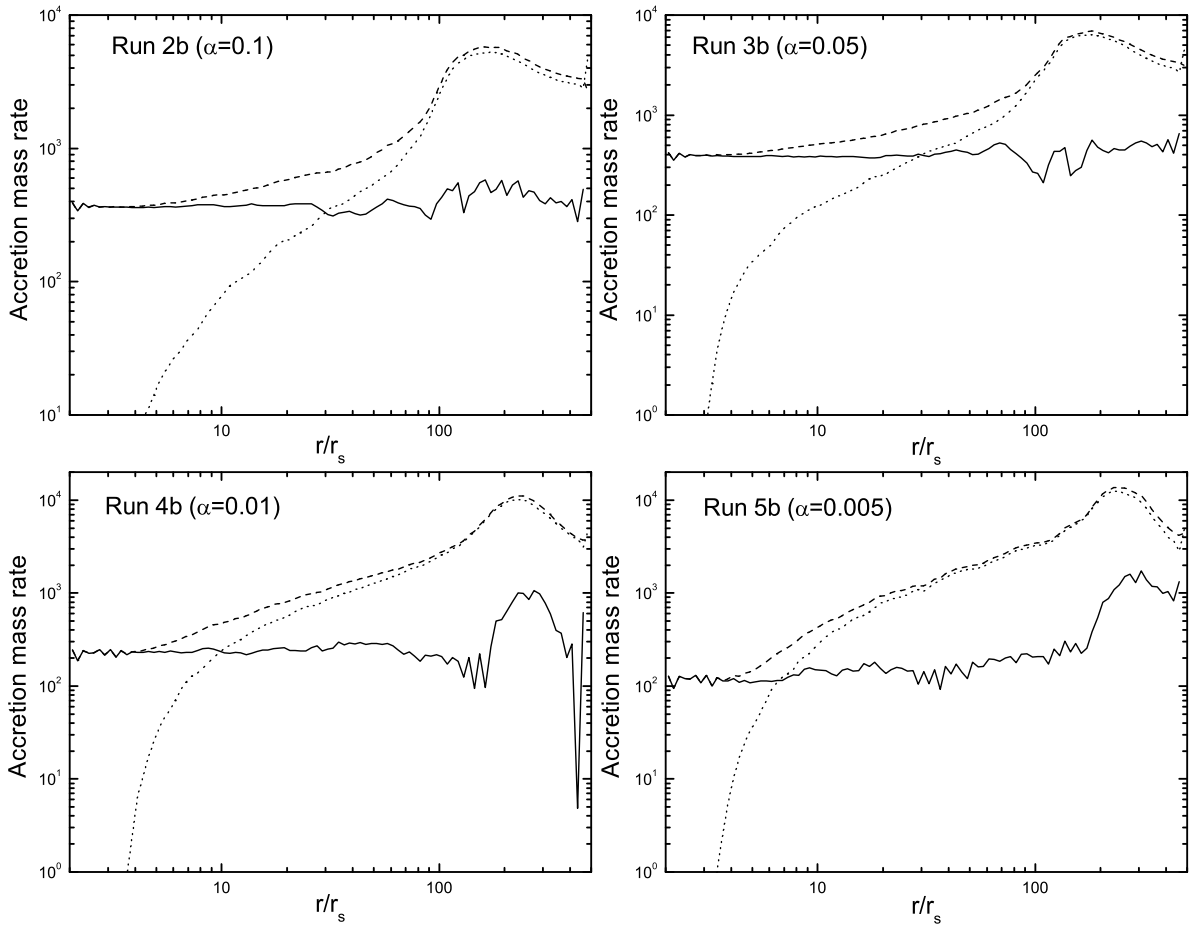}}}

\centering \caption{Radial profiles of inflow (dashed line),
outflow (dotted line), and net rates (solid line) of Model B for
various $\alpha$. This result is very similar to the case of hot
accretion flows, see text for details.}\label{fig 8}
\end{figure*}

\begin{figure*}

\scalebox{0.65}[0.65]{\rotatebox{0}{\includegraphics[bb=37 26 398
290]{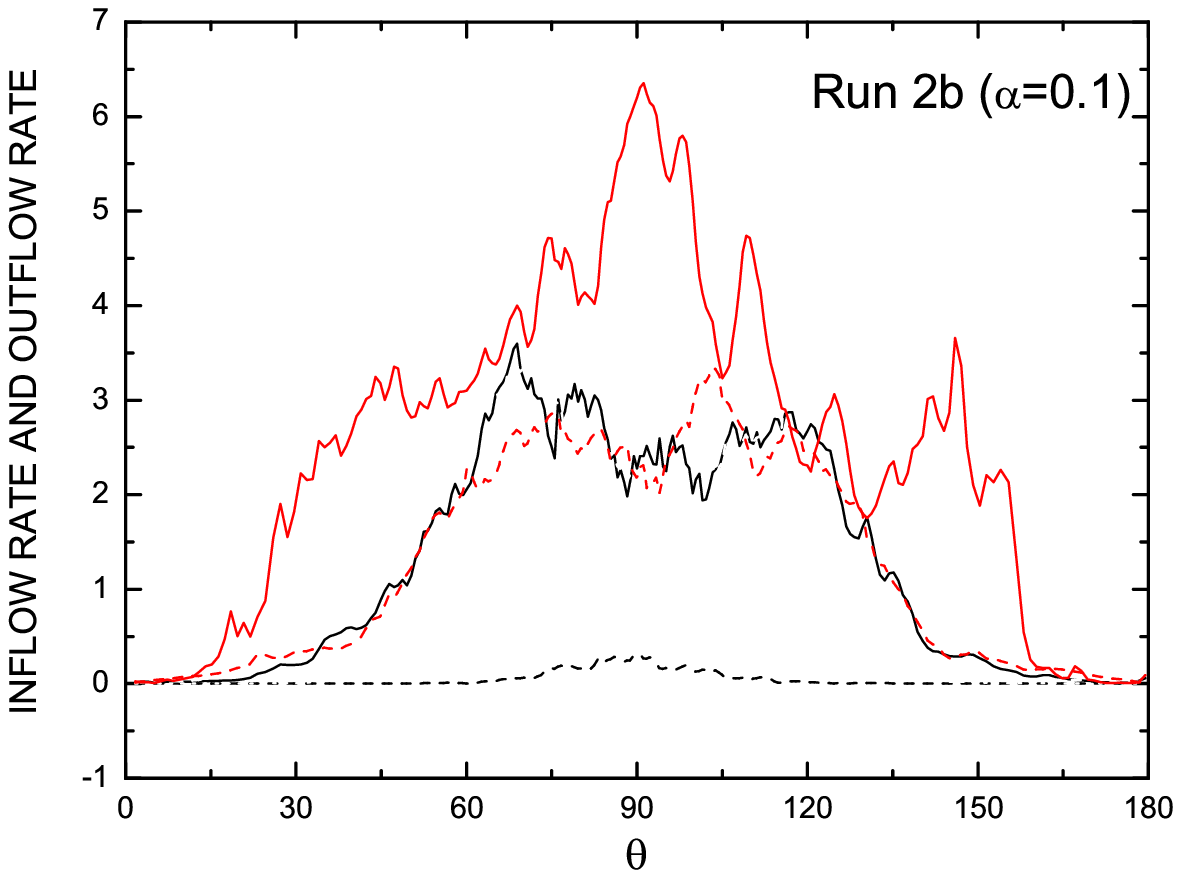}}}
\scalebox{0.65}[0.65]{\rotatebox{0}{\includegraphics[bb=37 26 398
290]{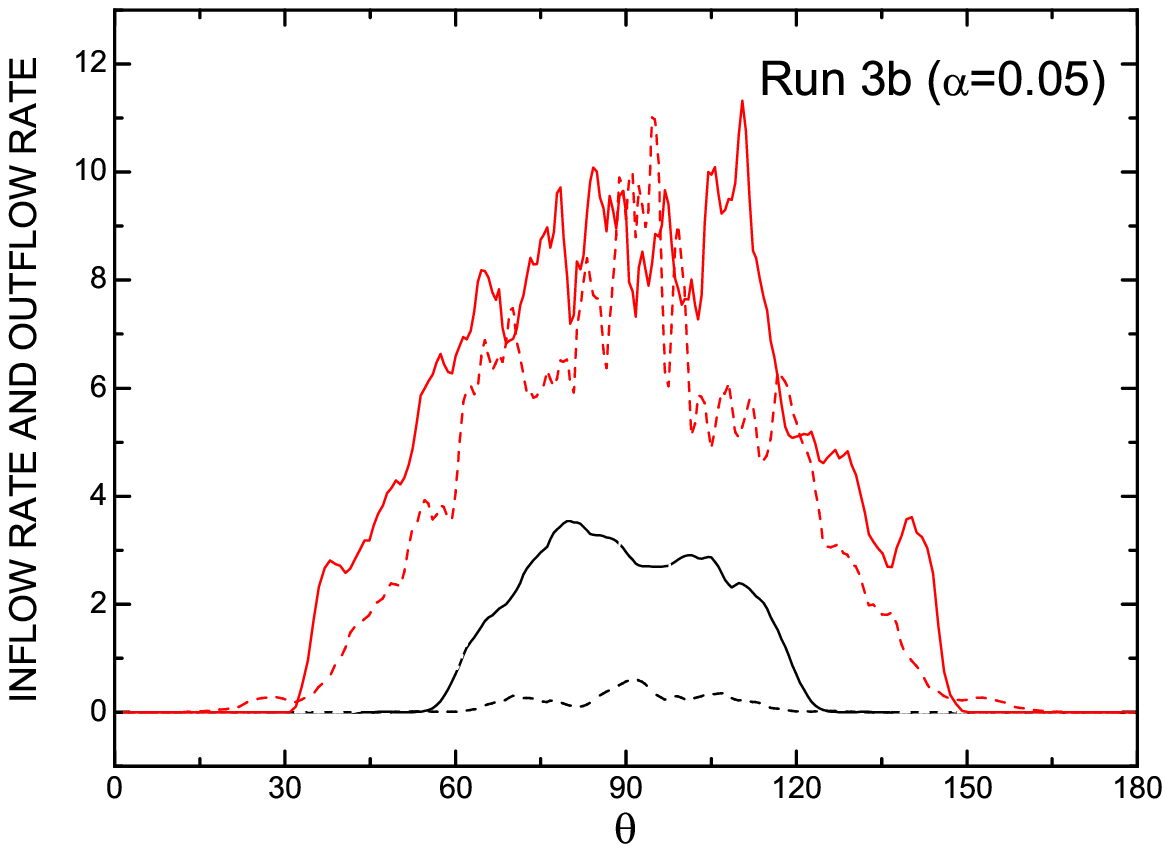}}}

\scalebox{0.65}[0.65]{\rotatebox{0}{\includegraphics[bb=37 26 398
290]{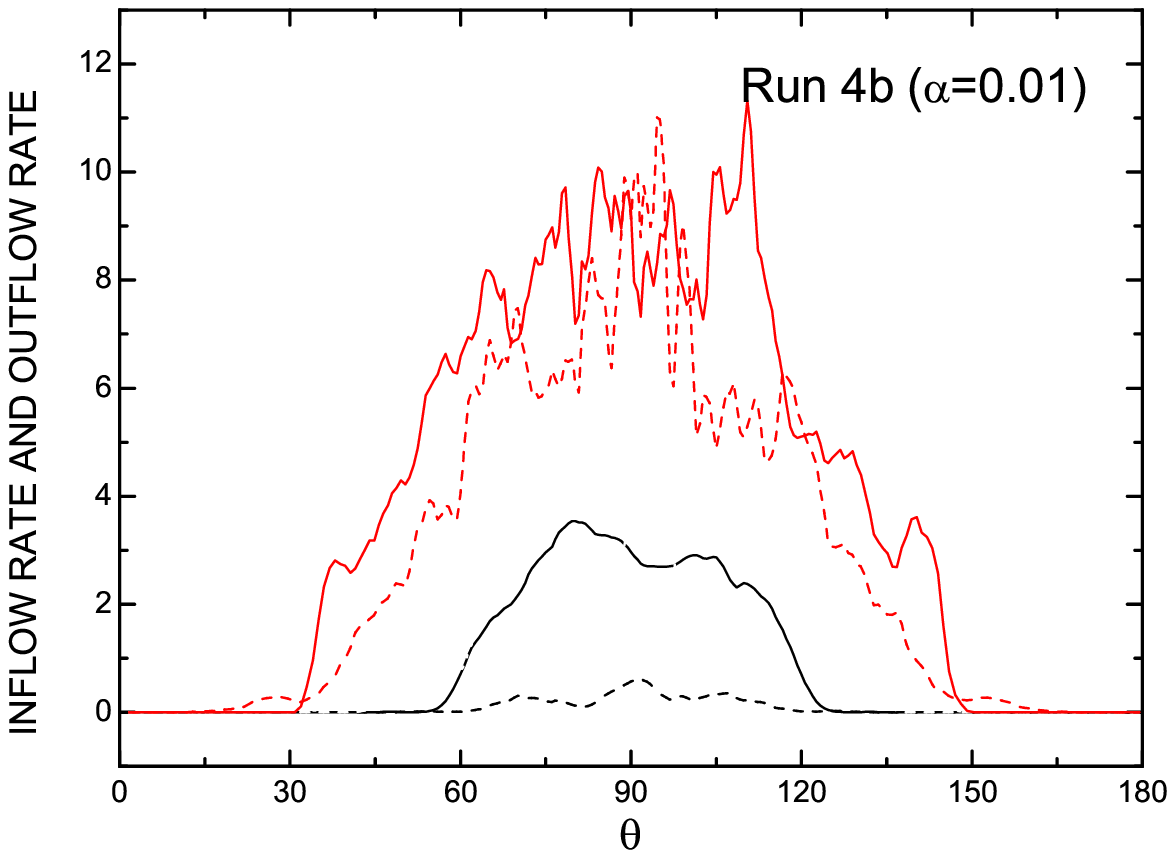}}}
\scalebox{0.65}[0.65]{\rotatebox{0}{\includegraphics[bb=37 26 398
290]{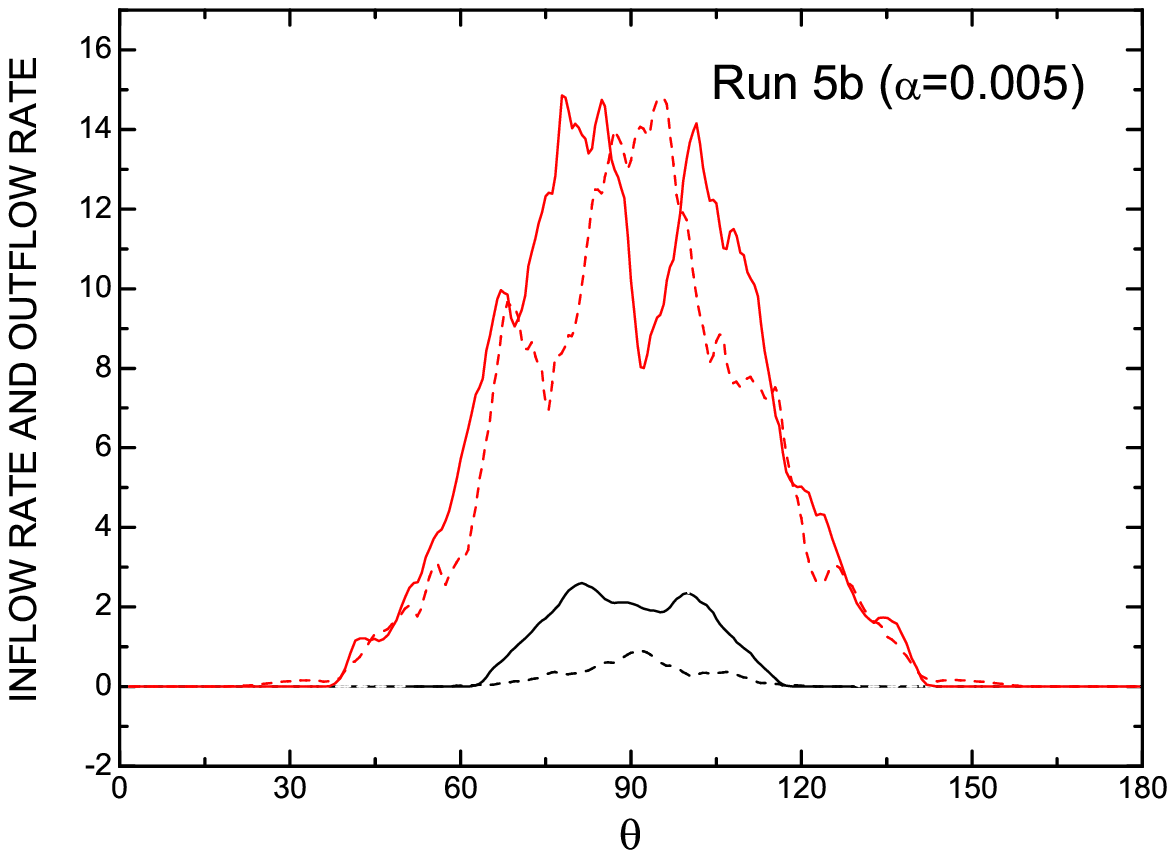}}}

\centering \caption{Angular distribution of inflow (solid lines) and
outflow (dashed lines) rates of Model B. The black and red lines
are for $r=5r_s$ and $r=30r_s$. } \label{fig 9}
\end{figure*}

In order to understand the angular profiles of radial velocity, we
plot in Figuer. \ref{fig 4} the angular distribution of radial
forces acting on unit mass for Run 5b at $r=5$ (top panel) and $30
r_s$ (bottom panel), respectively. We can see that the angular
distribution of net force (solid lines) is similar to the angular
distribution of $v_{r}$ shown in Figure \ref{fig 3}. In addition,
Figure \ref{fig 4} shows that within $10^{\circ}<\theta<90^{\circ}$
the radial component of centrifugal force efficiently counteracts
the gravity and even exceeds the gravity at some degree. The top
panel shows that the flows within $30^{\circ}$ from the equator are
super-Keplerian at $30 r_s$. At $r\sim10 r_s$, the flows are
super-Keplerian within $40^{\circ}$ from the equator. However,
within $\sim20^{\circ}$ from the axis, the radial component of
centrifugal force rapidly decreases and cannot efficiently
counteract the gravity. Hence, the equivalent potential well is
deeper near the axis than near the equator. On the other hand, the
radial component of the gas-pressure gradient force is negligible
within $60^{\circ}$ of the equator, while near the axis this
component is not negligible but is not the dominant force. The
radial net force within $60^{\circ}$ of the equator is dominated by
the radial radiation force, the gravity, and the radial centrifugal
force. In the inner region, the radial net force near the axis is
dominated by the gravity, so that we see that there is inflow near
the axis. In the outer region, the radial net force near the axis is
dominated by the radial radiation force, so that we can see that
there is strong outflow near the axis. Hence, the radial radiation
force plays an important role in maintaining the radial equilibrium
of flows near the equator and driving the high-latitude outflow in
the outer region.

Figure \ref{fig 5} shows the angular distribution of angular forces.
It is seen that the angular component of net force is nearly zero
near the equator, which indicates that flows are in force
equilibrium near the equator. This is because that the angular
component of gas-pressure gradient force is also negligible, and the
radiation force balances the centrifugal force in the
$\theta$-direction. The angular motion of flows near the axis is
controlled by the centrifugal force and gas-pressure gradient force,
while the angular component of radiation force is negligible near
the axis.

\begin{figure*}

\scalebox{0.65}[0.65]{\rotatebox{0}{\includegraphics[bb=36 23 379
275]{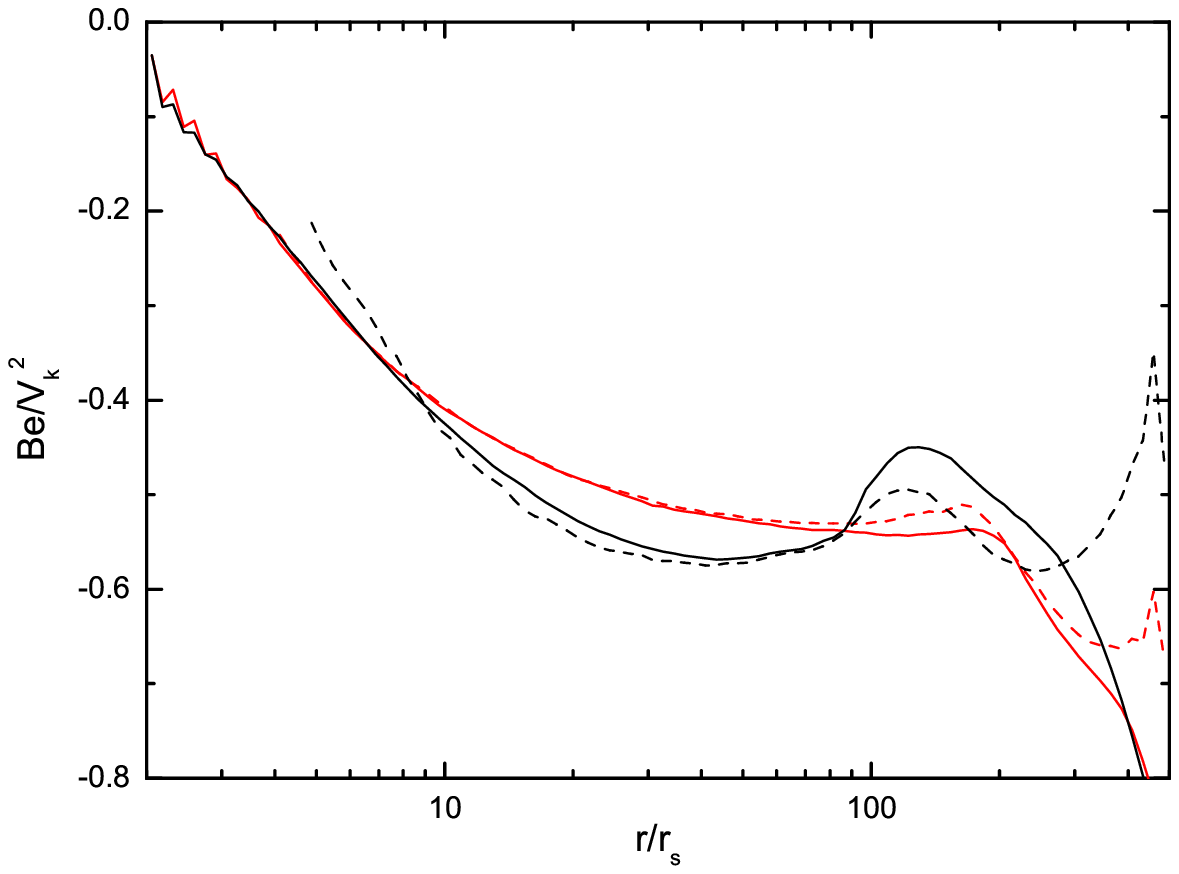}}}
\scalebox{0.65}[0.65]{\rotatebox{0}{\includegraphics[bb=36 23 379
275]{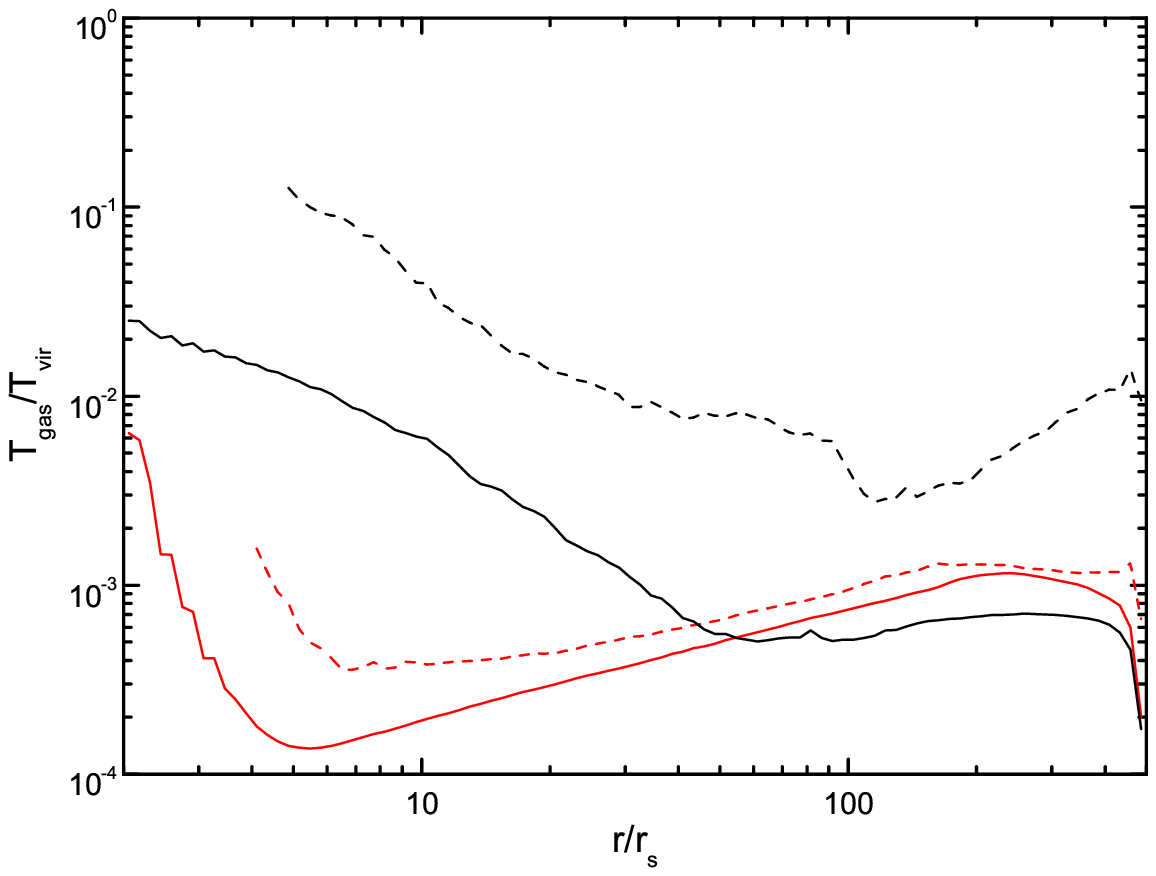}}}

\scalebox{0.65}[0.65]{\rotatebox{0}{\includegraphics[bb=36 23 379
275]{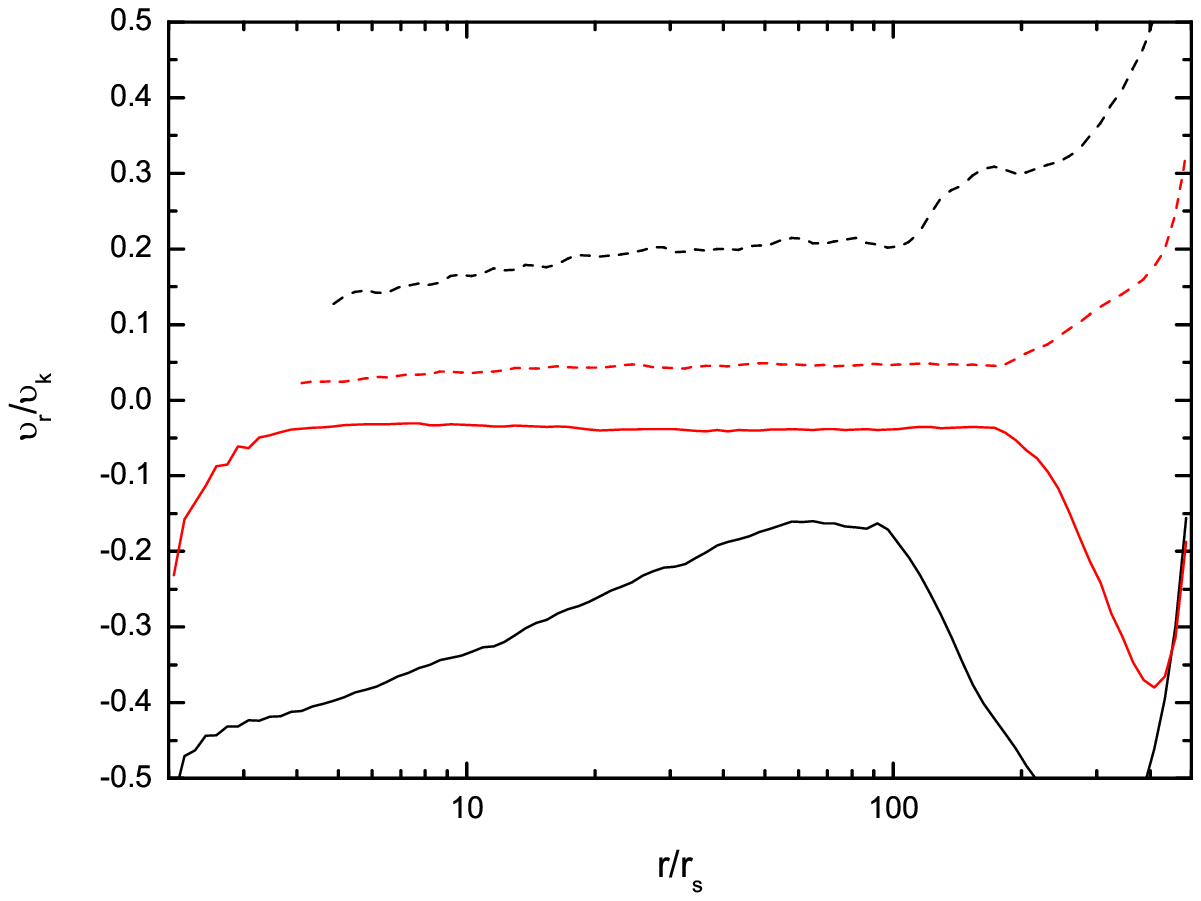}}}
\scalebox{0.65}[0.65]{\rotatebox{0}{\includegraphics[bb=36 23 379
275]{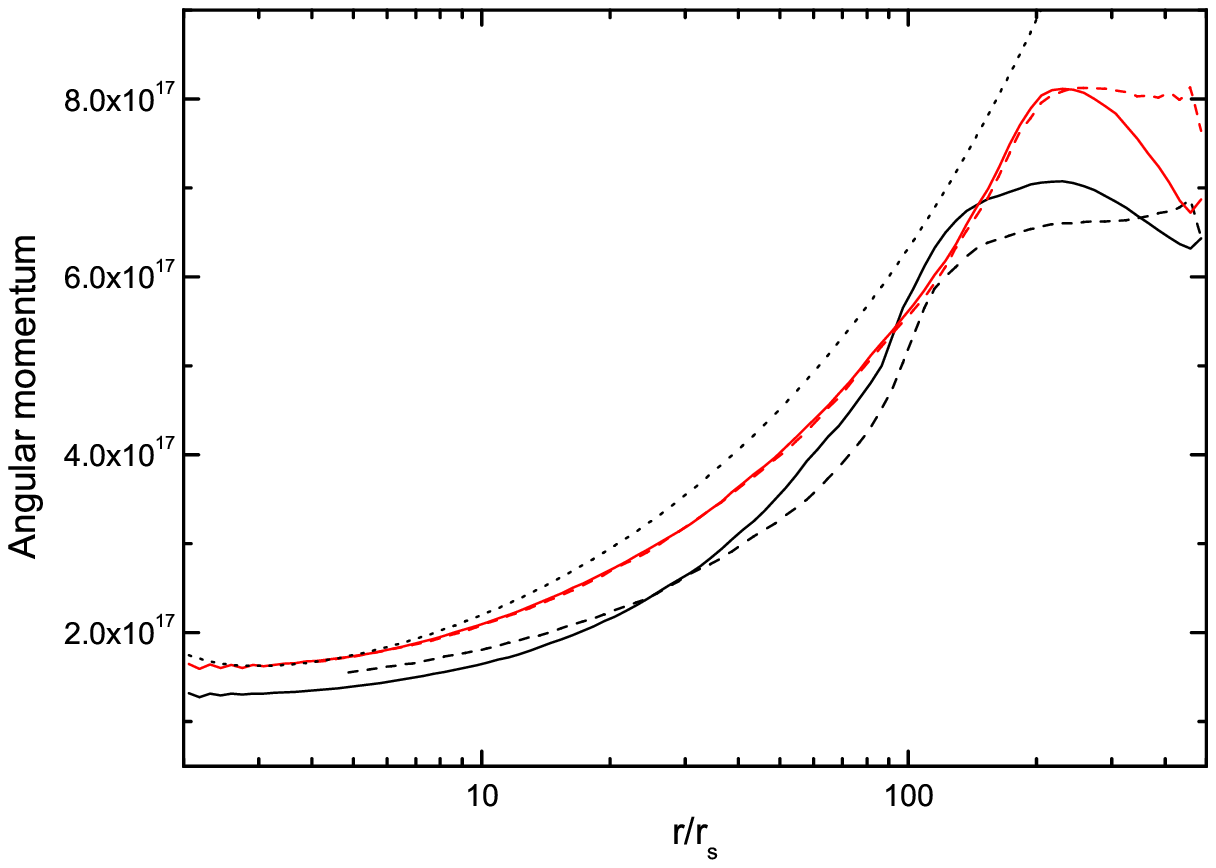}}}

\centering \caption{Radial distribution of time-averaged values
of flux-weighted Be (in units of $v_k^2$), $T_{\rm gas}$ (in units
of  $T_{\rm vir}$), $v_r$ (in unit of $v_k$), and specific angular
momentum. The solid and dashed lines are for inflow and outflow,
respectively. The black and red lines are for Run 2b ($\alpha=0.1$)
and Run 5b ($\alpha=0.005$), respectively. The dotted line in the
bottom right panel denotes the Keplerian angular momentum at the
equator. } \label{fig 10}
\end{figure*}

\subsection{Radial Structure of Accretion Flows}

The radial structure of a slim disk has been solved using a
vertical-integrated one-dimensional method (e.g., Abramowicz et al.
1988). However, the solution is one dimensional, thus the method
cannot treat more viscous components besides $T_{r\phi}$ and cannot
address outflow. Here, on the basis of the solution of
two-dimensional simulation, we plot the radial structure of the
time-averaged flow near the equator in Figure \ref{fig 6}. The
solution is averaged over the angle between $\theta =84^{\circ}$ and
$\theta =96^{\circ}$. In Figure \ref{fig 6}, the left panel is for
Run 2b (red line) and Run 3b (blue line), while the right panel is
for Run 4b (red line) and Run 5b (blue line).

Panel (A) of Figure \ref{fig 6} shows the radial profiles of density.
The dot-dashed lines indicate the self-similar solution of the
slim-disk model (Wang \& Zhou 1999). We find that the density
profile of different models can be well described by a power law
function, $\rho(r)\propto r^{-p}$ with $p\approx 0.55$. This result
is much flatter than the self-similar solution of the slim disk
where $\rho(r)\propto r^{-1.5}$ (Wang \& Zhou 1999), but it is very
similar to the case of hot accretion flows (Stone et al. 1999; Yuan
et al. 2012, and references therein). The reason for the discrepancy
is that in the self-similar solution the mass accretion rate
$\dot{M}(r)$ is assumed to be a constant of radius, while as we will
show in Section 3.4, the accretion rate actually decreases inward
because of the mass loss in outflows. Moreover, it is interesting to
note that the value of power law index of the density profile, $p$,
is quite ``universal'' for different models, although these models
have different $\alpha$ and different radial profiles of inflow rate
(refer to Section\ref{outflow}). Bu et al. (2013) studied the
effects of initial and boundary conditions in simulations of
accretion flow. They find a similar result, namely the density
profile is more converged compared with the diverse radial profile
of inflow rate.

Our result is apparently different from Figure 11 in Ohsuga et al.
(2005), in which they find $\rho(r)\propto r^{-1.5}$. For comparison
with Ohsuga et al. (2005), we have checked the models in Ohsuga et
al. (2005) and found that only when $\dot{m}_{\rm input}=500$ and
1000, and $\alpha=0.1$, the density profile can be described by
$\rho(r)\propto r^{-1.5}$. All other models have $\rho(r)\propto
r^{-0.55}$. Moreover, in the former case, the $\rho(r)\propto
r^{-1.5}$ density profile only holds in the range of $r\la 30 r_s$.
Beyond this radius, the profile becomes much flatter. But in the
latter case, the $\rho(r)\propto r^{-0.55}$ density profile holds
until $r\sim 80 r_s$, as shown by Figure 6. The reason for the
discrepancy is unclear.

\begin{figure*}

\scalebox{0.5}[0.5]{\rotatebox{0}{\includegraphics[bb=32 360 485
668]{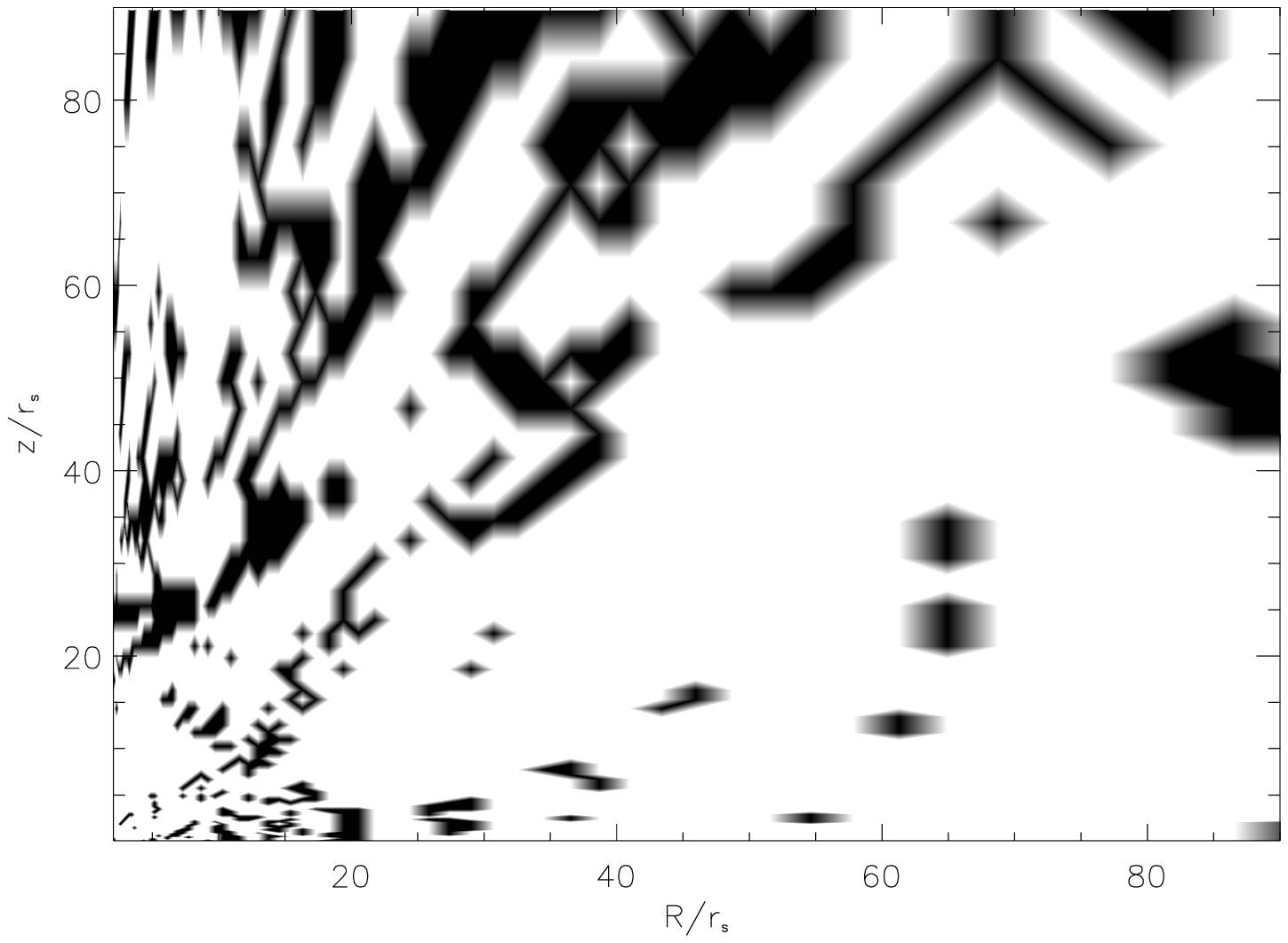}}}
\scalebox{0.5}[0.5]{\rotatebox{0}{\includegraphics[bb=32 360 485
668]{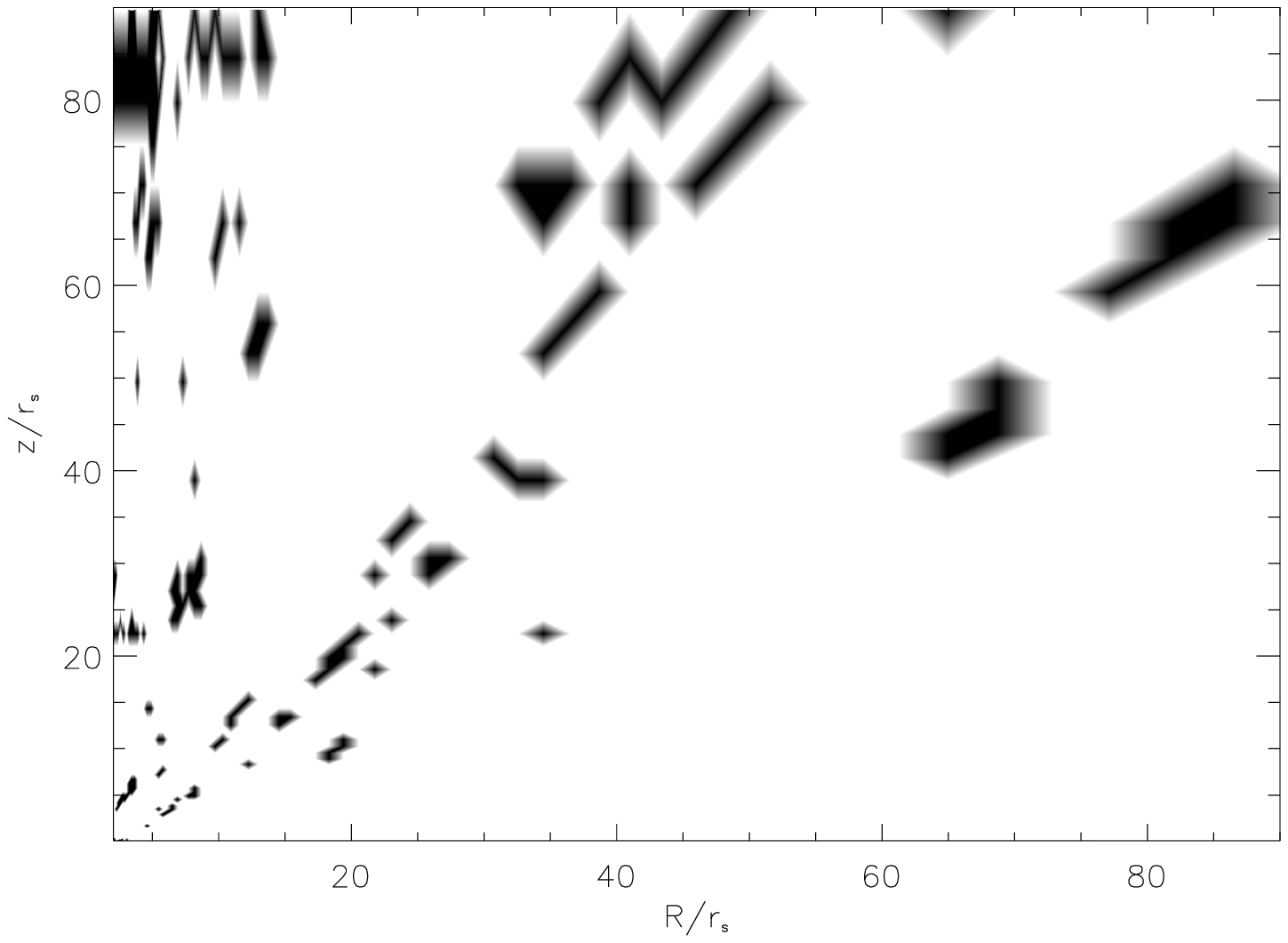}}}


\centering \caption{Convective stability analysis of Run 2b
($\alpha=0.1$; left) and Run 5b ($\alpha=0.005$; right) at $t=69.233 $
and $230.776$ orbits, respectively. The dark region denotes negative
$N_{\rm eff}^2$, i.e., convectively unstable region.} \label{fig 11}

\end{figure*}

Panel (B) of Figure \ref{fig 6} shows the radial profiles of gas
temperature $T_{\rm gas}$ (red and blue dashed lines) and
``radiation'' temperature $T_{\rm rad}$ (red and blue solid lines).
$T_{\rm rad}$ can be approximately described by a radial power law
function of $T_{{\rm rad}} \propto r^{-0.42}$, as the black solid
line shows. This is again flatter than the self-similar solution of
the slim-disk model, which has $T_{{\rm rad}} \propto r^{-5/8}$ as
shown by the dot-dashed line. We think the reason is because the
compression work becomes weaker because of the presence of outflow.
In the inner region close to the black hole, $T_{{\rm rad}}$ becomes
flatter. We can find that $T_{{\rm gas}}$ and $T_{{\rm rad}}$ are
nearly equal in the outer region, while the gas temperature is
higher than $T_{{\rm rad}}$ in the inner region. The radius where
the two temperatures deviate depends on the mass-injection rate and
viscous parameter $\alpha$. The higher the mass-injected rate and
the smaller the $\alpha$, the smaller the ``deviation'' radius.
However, when $\alpha=0.01$ and 0.005, the ``deviation radius'' is
located around the inner boundary. The discrepancy of the two
temperatures is because of the inefficient coupling between gas and
radiation. The energy transfer between gas and radiation is
controlled by the absorption opacity $\kappa_\text{p}$ (refer to the
term $|4\pi \kappa_\text{p} B-c \kappa_\text{p} E_0|$ ($B$ is the
blackbody radiation intensity) of Equations (7) and (8) in Ohsuga et
al. 2005). The absorption opacity $\kappa_\text{p} \propto \rho^2
T_\text{gas}^{-7/2}$ is due to free-free absorption and bound-free
absorption. We find $\kappa_\text{p} \propto r^{0.37}$, i.e.,
$\kappa_\text{p}$ decreases inward. Therefore, at a small radius,
the coupling between radiation and gas is weak, and the temperature
equilibrium between the gas and the radiation field is not achieved
before the gas falls onto the black hole. Smaller $\alpha$ gives
rise to smaller radial velocity, which provides more time to
transfer the energy of the gas to the radiation field. So, the
``deviation radius'' moves inward with the decrease of $\alpha$.

Panel (D) in Figure \ref{fig 6} shows the radial distribution of the
specific angular momentum. We can see that $\alpha$ can affect the
angular momentum distribution, especially in the vicinity of the
black hole ($r<$10$r_s$). In the case of $\alpha$=0.01 and 0.005,
the specific angular momentum becomes super-Keplerian in the range
of 3$-$6 $r_{s}$. $\dot{m}_{\rm input}$ also affects the angular
momentum distribution. The models with smaller $\alpha$ and lower
$\dot{m}_{\rm input}$ have slightly flatter distribution. This
result does not agree with that of Abramowicz et al. (1988).
Abramowicz et al. (1988) identified the tendency of the
specific angular momentum distribution to become flatter with the
increase of the accretion rate when the accretion rate is less than
$800\dot{M}_{\rm{crit}}$. They could not study the higher accretion
rate, because their method fails for
$\dot{M}>800\dot{M}_{\rm{crit}}$. The reason for the discrepancy is
unclear.

To analyze the energetics of all the models, we define the advection
factor ($q_{\text{adv}}$) of accretion flow as follows:
\begin{equation}
q_{\rm adv}\equiv\frac{Q_{\rm adv}}{Q_{\rm vis}}=1-\frac{Q^{-}_{\rm
rad}}{Q_{\rm vis}},
\end{equation}
where $Q_{\rm adv}$ is the gas and radiation energy advection rate,
$Q_{\rm vis} = T^2/\eta$ is the viscous dissipation rate, and
$Q^{-}_{\rm rad}$ is the radiation cooling rate. Here $Q^{-}_{\rm
rad}$ is defined as the $\theta$ component of $-\nabla \cdot F$ ($F$
is the radiation flux). Both $Q^{-}_{\text{rad}}$ and $Q_{\rm{vis}}$
are obtained first by time-averaging and then by averaging the
quantity over an angle between $\theta =84^{\circ}$ and $\theta
=96^{\circ}$. Figure \ref{fig 7} shows the time-averaged advection
factor near the equator for the models of
$\dot{m}_{\rm{input}}=3000$ (i.e., Runs 2b, 3b, 4b, and 5b). We can
see that $q_{\rm{adv}}$ is approximately close to a constant in the
range of $(5-70) r_s$. The value of $q_{\rm{adv}}$ deceases inward
when $r<10 r_s$, especially for the models with small $\alpha$.
Since the opacity of accretion flow is dominated by scattering
opacity and the half-thickness of accretion decreases inward, the
vertical opacity of accretion flow decreases inward. Therefore, the
photon trapping effect becomes weaker and the radiation becomes
stronger. $Q^{-}_{\rm{rad}}$ can even be larger than $Q_{\rm{vis}}$,
so $q_{\rm{adv}}$ is negative, as shown by Figure \ref{fig 7}. This
is consistent with Abramowicz et al. (1988). This indicates that
advection plays a heating rather than cooling role, similar to the
case of luminous hot accretion flows (Yuan 2001).

\begin{figure*}
\scalebox{0.5}[0.5]{\rotatebox{0}{\includegraphics[bb=58 42 532
386]{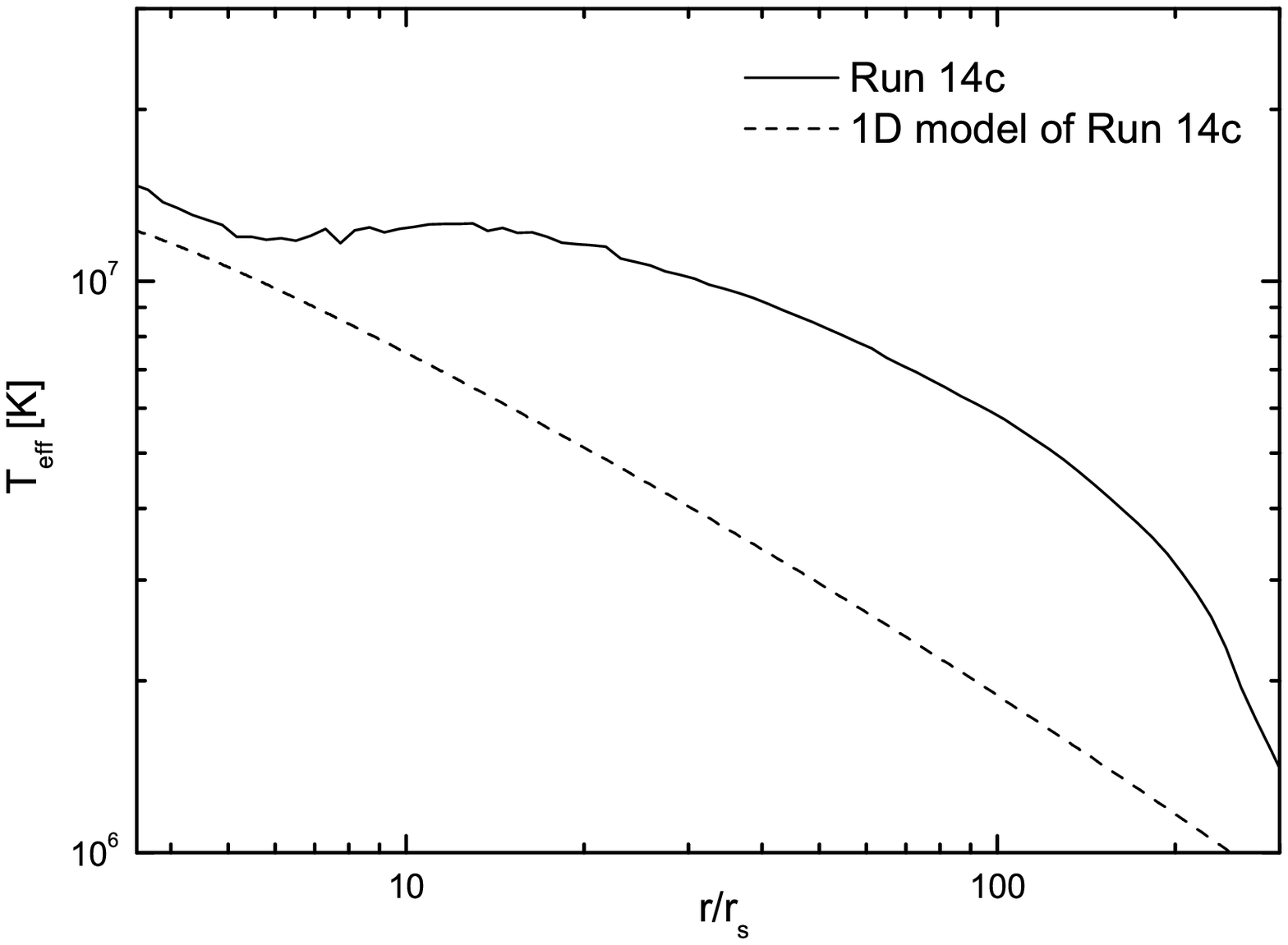}}}
\scalebox{0.5}[0.5]{\rotatebox{0}{\includegraphics[bb=58 42 584
386]{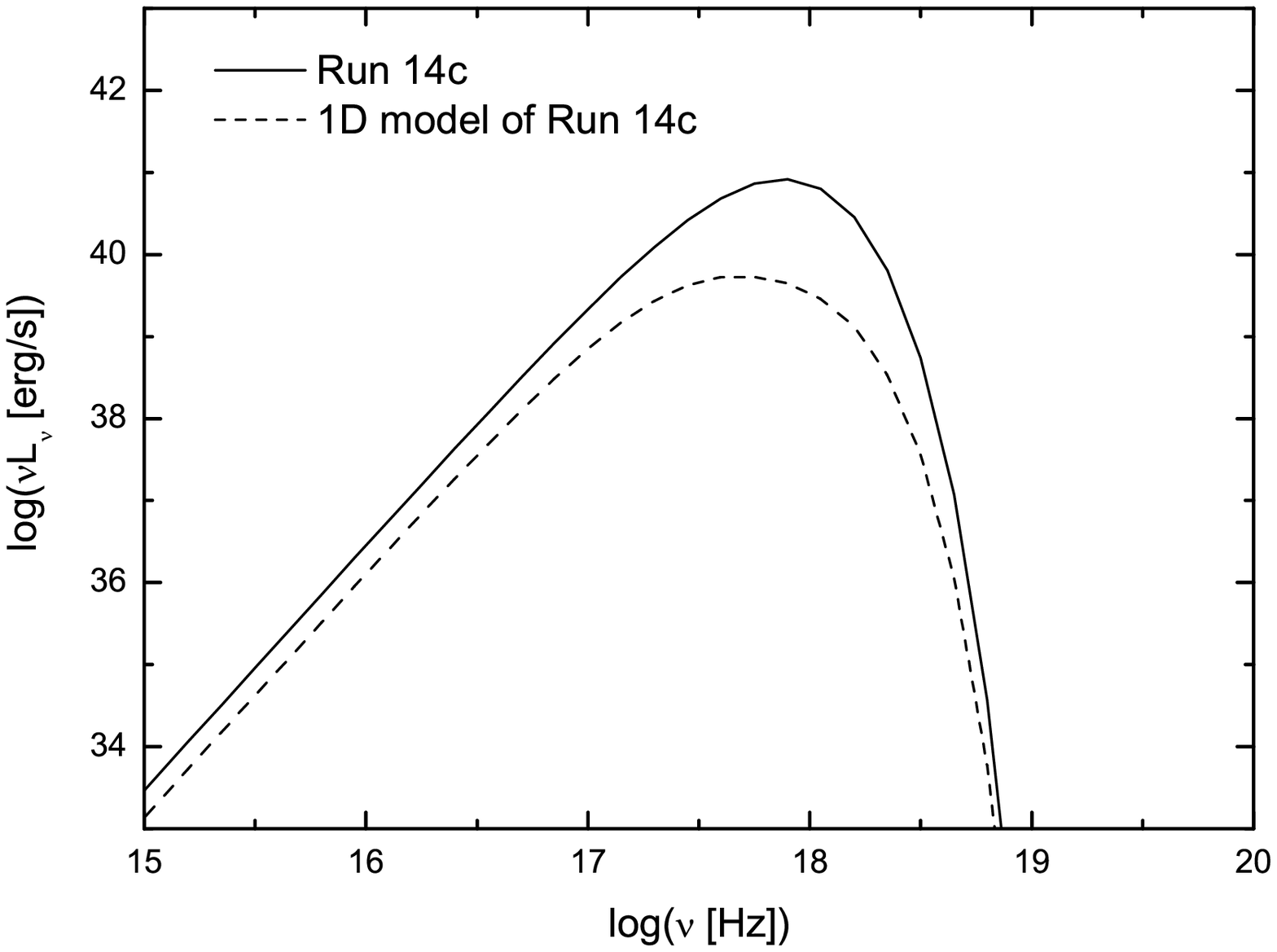}}} \vspace{0.5cm} \centering \caption{Effective temperature (left) and multi-color blackbody spectra
(right) of a slim-disk model (Run 5b with $\alpha=0.005$ and
$T_{r\phi}$ and $T_{\theta\phi}$). The solid and dashed lines
correspond to the numerical simulation and the one-dimensional global
solution, respectively.} \label{fig 12}
\end{figure*}

\subsection{Inflow and Outflow: Rates and Properties}
\label{outflow}

Following Stone et al. (1999), we define the mass
inflow ($\dot{m}_{\rm in}$) and outflow rates
($\dot{m}_{\rm{out}}$), in units of the critical mass accretion rate
$\dot{M}_{\rm {crit}}$, as the following time-averaged and
angle-integrated quantities:
\begin{equation}
\dot{m}_{\rm in}(r)= -\frac{c^2}{L_{\rm Edd}}\int^{\pi}_{0} 2\pi
r^2\rho \text{min}(\bm{v}_r,0)\text{sin}\theta d\theta,
\label{inflowrate}
\end{equation}
\begin{equation}
\dot{m}_{\rm out}(r)=\frac{c^2}{L_{\rm Edd}}\int^{\pi}_{0} 2\pi r^2
\rho \text{max}(\bm{v}_r,0)\text{sin}\theta d\theta.
\label{outflowrate}
\end{equation}
The net mass accretion rate,
$\dot{m}_{\rm net}(r)=\dot{m}_{\rm in}(r)-\dot{m}_{\rm out}(r)$, is the
accretion rate that finally falls onto the black hole. It is noted
that the above rates are obtained by time-averaging the integral rather
than integrating the time-averages. According to Equation (5), we know that
the outflow rate is not a correct measurement of real outflow,
although it can provide some important information. This is because
it includes the contribution of turbulent motion.

Figure \ref{fig 8} shows the radial distribution of inflow, outflow,
and net rates of Model B ($\dot{m}_{\rm {input}}=3000$; i.e., Runs
2b, 3b, 4b, and 5b). We can see that the inflow and outflow rates
decrease inward, as in the case of hot accretion flows (Stone,
Pringle et al. 1999; Yuan et al. 2012, and references therein). The
radial profiles of $\dot{m}_\text{in}$ and $\dot{m}_\text{out}$ can
be described by a power law function of radius. Using
$\dot{m}_\text{in}(r)\propto r^s$ to fit the inflow rate in the
range of $(8-50)r_s$, we find that the value of $s$ is not sensitive
to $\dot{m}_{\rm input}$  but mainly determined by the value of
$\alpha$. A large $\alpha$ corresponds to a small $s$.  We have
calculated models with different $\dot{m}_{\rm{input}}$ and derived
the average values of $s$. They are $\sim0.37, 0.44, 0.76$ ,and
$0.98$ for models with $\alpha=0.1, 0.05, 0.01$, and $0.005$,
respectively. The dependence of $s$ on $\alpha$ can be approximately
described by a power law function $s\propto \alpha^{-0.33}$. This
result is again similar to the case of hot accretion flows. Yuan et
al. (2012) find that $s\sim 0.54$ and $0.65$ for $\alpha=0.01$ and
$0.001$, respectively. Note that the definition of $\alpha$ is
different in the two works. We find that the value of $\alpha$ in
this paper corresponds to a smaller $\alpha$ in Yuan et al. (2012).
The quantitative comparison of the value of $\alpha$ is difficult.
As argued in Yuan et al. (2012) and F. Yuan et al. (2013, in
preparation), in the case of hot accretion flow, the inward decrease
of inflow rate is due to mass loss via outflows. We believe this is
also the case for the slim disk. The similar slope in the slim disk
and hot accretion flow implies that the strength of outflow is
similar in the two cases. We will discuss the possible origin of
outflow below.

In order to analyze the angular distribution of the mass accretion
rate, we time-average the following inflow and outflow rates as a
function of $\theta$:
\begin{equation}
\dot{m}_{\rm in}(\theta)=  -2\pi r^2\rho
 \text{min}(\bm{v}_r,0)\text{sin}\theta \Delta\theta \frac{c^2}{L_{\rm Edd}},
\end{equation}
\begin{equation}
\dot{m}_{\rm out}(\theta)=2\pi r^2 \rho
 \text{max}(\bm{v}_r,0)\text{sin}\theta \Delta\theta \frac{c^2}{L_{\rm Edd}}.
\end{equation}
Similar to Equation (5), Equation (7) also does not correctly
measure the angular distribution  of real outflow. The results are
shown in Figure \ref{fig 9} for Model B. We can see that their
angular distributions are nearly symmetric to the equator and become
broader with the increase of radius and/or $\alpha$. The
distribution is different than the case of hot accretion flows
(refer to Figures 2 and 3 in Yuan et al. 2012). Firstly, in the case
of slim disks, the inflow and outflow rates are ``synchronous'',
i.e, they reach their maximum at the same $\theta$ angle. But in the
case of hot accretion flows, they are not (compare Figure \ref{fig
9} in the present paper and Figures 2 and 3 in Yuan et al. 2012).
Secondly, the outflow rate centers around the equatorial plane in
the present case, while in the case of hot accretion flows, the
maximum of outflow rate is located roughly at the surface of the
disk (their Model B and Model C). In that case, the inflow and
outflow rates are not symmetric to the equatorial plane. The reason
for the discrepancy is unclear. The outflow becomes stronger with
the increase of radius. This is consistent with Figure \ref{fig 8}
and is easy to understand. Compared with the region around the
equator, the inflow and outflow rates are almost negligible at the
high-latitude region, where the radial velocity of outflow is very
high. This is because the density of high-latitude flows is very
low. The high-latitude outflow is likely driven by radiation, but
its contribution to outflow rate seems to be negligible. The outflow
rate is dominated by the low-latitude outflow. The nature of the
low-latitude outflow, namely whether they are real systematic
outflow or simply turbulence, is an important question and needs to
be studied in the future. Another question is whether the origins of
outflow are real. In the case of hot accretion flows, Yuan et al.
(2012) identify the mechanism of producing the outflow by buoyancy
when magnetic field is absent. This is because a hydro accretion
flow is convectively unstable. In Section 3.5 of the present paper,
we analyze the convective stability of slim disks.

Following Yuan et al. (2012), we analyze some properties of inflow
and outflow, such as the Bernoulli parameter, gas temperature,
radial velocity, and angular momentum. The motivation is to study
the mechanism of producing outflow. We calculate the flux-weighted
quantities (refer to Equations (8) and (9) in Yuan et al. 2012) and
then time-average the quantities. Figure \ref{fig 10} shows the
radial distribution of flux-weighted quantities ($Be$ in units of
$v_k^2$, where $v_k$ is the Keplerian velocity; $T_{\rm gas}$ in
units of the virial temperature $T_{\rm vir}\equiv
\frac{GMm_p}{3kr}$, where $m_p$ is the photon mass and $k$ the
Boltzmann constant;and $v_r$ in unit of $v_k$) of inflow and outflow
for Run 2b and Run 5b, respectively. In both models,  Be is
negative. The value of Be is in general smaller than that of hot
accretion flows (Yuan et al. 2012), which is of course because the
energy loss in slim disk is stronger.

The top right panel in Figure \ref{fig 10} shows that the
temperature of outflow is higher than that of inflow. This seems to
suggest that the mechanism of outflow production is because of
buoyancy, like in the case of hydrodynamical hot accretion flow.
Moreover, the discrepancy of the two temperatures is larger when
$\alpha$ is larger (Run 2b). The convective stability of slim disk
is analyzed in Section \ref{convection}. The bottom left panel in
Figure \ref{fig 10} shows that the radial velocity of outflow can be
well described  by $v_r/v_k \sim$ const. This is similar to hot
accretion flows (Yuan et al. 2012). It is interesting to note that
when $\alpha$ is smaller, the radial velocities of both inflow and
outflow are smaller. For inflow, this is easy to understand. For
outflow, the discrepancy of the radial velocity may be related to
the discrepancy of the temperature between inflow and outflow. As we
can see from the top right panel, when $\alpha$ is smaller, the
discrepancy is smaller, and thus the buoyancy may be weaker. The
reason why the temperature discrepancy in the case of small $\alpha$
is smaller is perhaps related to the magnitude of the convective
energy flux. The convective energy flux transports energy along
surfaces of constant $r\sin^2(\theta)$ (Quataert \& Gruzinov 2000).
The magnitude of this flux is proportional to the convective
diffusion coefficient $\alpha_c$, and a smaller $\alpha$ corresponds
to a smaller $\alpha_c$ (Narayan et al. 2000). Therefore, when
$\alpha$ is larger, the fluids at the surface are more strongly
heated by the stronger convective energy flux. At last, the bottom
right panel shows that the angular momenta of inflow and outflow are
almost identical. This is similar to the case of hydrodynamical hot
accretion flows (Yuan et al. 2012).

\subsection{Convective Stability}
\label{convection}

In the case of hydrodynamical hot accretion flow, it is shown that
the inward decrease of inflow rate is because of mass loss in
outflow produced by the buoyancy associated with convective
instability (Yuan et al. 2012). In this section, we analyze the
convective stability of the slim disk on the basis of our simulation data.

The energy equation of accretion flows can be written as
\begin{equation}
Q^-_{\rm adv}=Q^+_{\rm vis}-Q^-_{\rm rad},
\end{equation}
where $Q^-_{\rm adv}=\Sigma \upsilon_r T (dS/dr)$ ($\Sigma$ is
surface density, and $S$ is the specific entropy). In general,
$Q^+_{\rm vis}>Q^-_{\rm rad}$ so that $\Sigma \upsilon_r T
(dS/dr)>0$ and $dS/dr<0$. The inward increase of entropy is thought
of as a necessary condition of convective instability for rotating
flow. A series of simulation studies (Igumenshchev \& Abramowicz
1999; Stone et al. 1999; Igumenshchev \& Abramowicz 2000; Yuan \& Bu
2010) verified that hot accretion flows are convectively unstable,
confirming the prediction of Narayan \& Yi (1994). Compared with hot
accretion flows, the physics of a slim disk is different. In a slim
disk, the gas particles can efficiently radiate, but the photons are
trapped and hence cannot efficiently escape from the system because
of the large scattering optical depth. In addition, a slim disk is
supported by the radiation pressure, and hence the specific entropy
is dominated by the radiation photons. Sadowski et al. (2009, 2011)
found that a radiation pressure-supported disk is convectively
unstable. Gu (2012) revisited this problem, taking into account the
local energy balance between the viscous heating and the advective
and radiative cooling, and found that a slim disk is convectively
stable. Gu (2012) thought that the significant difference in the
results between their work and Sadowski et al (2009). is probably
related to the different approaches for describing the vertical
structure. But in Gu (2012), a self-similar solution of the radius
is adopted, the consequence of which is unknown.

Here we analyze the convective stability of a slim disk on the basis of
our simulation data. We assume that the H\"oiland criterion (e.g.,
Begelman \& Meier 1982) is applicable to the slim disk. The
condition in cylindrical coordinates ($R,\phi,z$) for convective
stability in a rotating accretion flow (Tassoul 1978) is
\begin{equation}
N^2_{\rm eff} = N^2_{R}+N^2_{z}+\kappa^2>0,
\end{equation}
where $N_{\rm eff}$ is defined as an effective frequency, $\kappa$
is the epicyclic frequency which is calculated by
$\kappa^2=\frac{1}{R^3}\frac{\partial l^2}{\partial R}$ ($l$ is the
specific angular momentum), and $N_{R}$ and $N_{z}$ is the $R$ and
$z$ component of the well-known Brunt--V\"{a}is\"{a}l\"{a}
frequency, respectively, which can be calculated by
\begin{equation}
N^2_{R}=-\frac{1}{\gamma_r \rho}\frac{\partial P}{\partial
R}\frac{\partial S}{\partial R}
\end{equation}
and
\begin{equation}
N^2_{z}=-\frac{1}{\gamma_r \rho}\frac{\partial P}{\partial
z}\frac{\partial S}{\partial z},
\end{equation}
where $dS \propto d\text{ln}(\frac{P}{\rho^{\gamma_r}})$, $P$ is the
total pressure, and $\gamma_r$ is the adiabatic index. Here $P=p_g$
and $\gamma_r=5/3$ are employed for the flows of low-density
($\rho<10^{-5}$ g/cm$^3$) and high-gas temperature ($T_{\rm gas} >
10^9$ K) (i.e., outflow or corona region), while $P=E_0/3$ and
$\gamma_r=4/3$ are employed for other radiation-dominated flows
(i.e., the disk body) where radiation and gas are effectively coupled
because of the large scattering opacity.

The results are shown in Figure \ref{fig 11}. The left panel is for
Run 2b at $t=69.233$ orbits, while the right panel for Run 5b at
$t=230.777$ orbits. The dark region denotes the unstable region,
i.e., $N^2_{\text{eff}}<0$. We can see that the results are somewhat
subtle. For Run 2b nearly half of the region is convectively
unstable, while for Run 5b the fraction of the unstable region is
much less. Given this result, we can perhaps say that Run 2b is
marginally convectively unstable and Run 5b is stable. In both
cases, compared with the case of a hydrodynamical hot accretion flow
(refer to Figure 6 in Yuan \& Bu 2010), the unstable region becomes
significantly smaller. We speculate that the difference between the
present work and a hydrodynamical hot accretion flow (i.e., Figure 6
in Yuan \& Bu 2010) is due to radiation. Recall that in an
magnetohydrdynamical hot accretion flow, the magnetic field plays a
role of stabilizing the convection (Balbus \& Hawley 2002; Narayan
et al. 2002). Here radiation plays an effectively similar role
although the underlying physics may be different. Physically, when
radiation is important as in a slim disk, radiation can directly
transport energy, thus the ``need'' for convection to transport
energy becomes weaker. Why is the unstable region in Run 2b  larger
than that in Run 5b? The only difference between Run 2b and Run 5b is
the value of $\alpha$, which is 0.1 and 0.005, respectively.
Compared with the case of a large $\alpha$, when $\alpha$ is smaller,
the convective energy flux is smaller, thus the role of radiation to
stabilize the flow becomes relatively more effective. This explains
the difference between Run 2b and Run 5b. This may also explain why the
temperature discrepancy between inflow and outflow in Run 2b is
larger than in Run 5b (refer to Figure \ref{fig 10}).

\subsection{Multi-color Blackbody Spectra}
\label{spectra}

As an initial step of the application of our numerical simulation,
in this section we compare the emitting spectra from two-dimensional
numerical simulations and one-dimensional analytical global
solutions. For simplicity, we assume a multi-color blackbody
spectrum on the basis of effective temperatures. In reality,
however, Compton scattering plays an important role in the emitted
spectrum (Kawashima et al. 2012). We identify the radiation
temperature $T_{\rm rad}$ at the photosphere (where the effective
optical depth equals 1) as the effective temperature. The effective
optical depth is calculated by
$\tau_{\textbf{eff}}=\sqrt{\tau_{\text{ab}}(\tau_{\text{ab}}+\tau_{\text{sc}})}$,
where $\tau_{\text{ab}}$ and $\tau_{\text{sc}}$ are the absorption
and Thomson scattering optical depth integrated from the outer
boundary along the \textit{Z}-direction. We adopt the free free
absorption opacity ($\kappa_{\text{ff}}$).

Taking Run 5b as an example, we have calculated the emitted spectrum
based on the simulation data. Only the range of $3.5 r_s<r<80 r_s$
is considered. The reason is that only the region of $r<80 r_s$ has
achieved steady state while the region of $r<3.5 r_s$ is effectively
optically thin due to small absorption optical depth caused by high
gas temperatures. We also calculate the global solutions of
one-dimensional model corresponding to the parameters of Run
5b(refer to Abramowicz et al. 1988 and Watarai et al. 2000 for the
calculation approach) and the corresponding spectrum. Figure
\ref{fig 12} shows the calculation results. We note that the
calculated luminosity based on our two-dimensional numerical
simulation agrees well with that obtained by Kawashima et al.
(2012). Significant differences of temperature and subsequently
emitted spectrum between the two models can be found. This result
can give us some initial idea of the difference between simple
one-dimensional calculation and the more realistic two-dimensional
simulation, which should be useful when applying the slim disk
theory to observations. The detailed calculation of the spectrum can
be found in Kawashima et al. (2012) and beyond the scope of the
present work.

\section{SUMMARY AND DISCUSSIONS}

In this paper, we have performed a two-dimensional radiative
hydrodynamical numerical simulation of slim disks. The technical
differences between the present work and Ohsuga et al. (2005) are
that we include an additional component of viscous stress, i.e.,
$T_{\theta\phi}$, and consider various values of the viscous parameter
$\alpha$. We find that the component $T_{\theta\phi}$ plays an
important role in transporting the angular momentum between different
latitudes. As a result, compared with the case of no $T_{\theta\phi}$
component (Ohsuga et al. 2005), the high-latitude outflow (within
$30^\circ$ from the axis) rotates slower, while the flow close to the
equatorial plane rotates faster. In addition, we find that the
high-latitude outflow has higher density, lower speed, and a smaller
Bernoulli parameter. For the effect of the magnitude of $\alpha$, we
find that the models with different $\alpha$ have similar radial
structure but different angular structure. The value of $\alpha$
strongly affects the radial velocity and the value of $Be$ of outflow at a
high latitude.

We have paid more attention in the present work to studying the
physics of slim disks. We have calculated the radial profiles of
inflow and outflow rates defined by Equations (4) and (5). We have found
that both of them decrease inward. Specifically, the inflow rate can
be well described by a power law form, $\dot{M}_{\rm in}\propto
r^{s}$. The value of $s$ is not sensitive to the accretion rate but is
mainly dependent on the value of $\alpha$. For $\alpha= 0.1, 0.05,
0.01$ ,and $0.005$, $s\sim 0.37, 0.44, 0.76$, and $0.98$, respectively
(Figure \ref{fig 8}). Correspondingly, the radial profile of density
becomes flatter compared with the case of a constant $\dot{M}(r)$. The
density profile can be described by $\rho(r)\propto r^{-p}$. It is
interesting to note that the value of $p$ is within a narrow range,
$p\approx 0.55$ for $\alpha\sim 0.005-0.1$ (Figure \ref{fig 6}).

These results are very similar to a hydrodynamical hot accretion
flow. In that case, Yuan et al. (2012) show that the inward decrease
of inflow rate is because of the mass loss in outflow. We believe
this is also the case for the present slim disk. In the case of hot
accretion flows, the mechanism of producing outflow is identified to
be buoyancy associated with the convective instability of the
accretion flow. To investigate the origin of outflow in the slim
disk, we first calculate and compare the properties of inflow and
outflow. We have found that the temperature of inflow is lower than
that of outflow. The discrepancy is larger when $\alpha$ is larger
(Figure \ref{fig 10}). This suggests the existence of convective
instability at some level, especially when $\alpha$ is large. We
then analyze the convective stability of the accretion flow on the
basis of our simulation data. The result is somewhat subtle. When
$\alpha=0.1$, about half of the region of the accretion flow is
convectively unstable, but when $\alpha=0.005$, less than half of
the region is unstable (Figure \ref{fig 11}). Recall that a
non-radiative accretion flow is convectively unstable, and we
speculate that radiation can stabilize the convection. Physically
this is because radiation can also take away energy, like
convection. The effectiveness of this stabilizing seems to depend on
the magnitude of $\alpha$. When $\alpha$ is smaller, it is more
effective, i.e., the accretion flow tends to be more convectively
stable (Section \ref{convection}).

Returning to the issue of the radial profile of inflow rate, two
questions arise. The first question is what is the mechanism of
producing outflow, especially if the slim disk is roughly
convectively stable when $\alpha$ is small? We speculate that the
outflow may be produced by radiation force. Or more precisely
speaking, both convection and radiation force can produce outflow.
Their relative importance may depend on $\alpha$. When $\alpha$ is
small, the convective energy flux is weaker, thus radiation force
will be the dominant mechanism of producing outflow. The force
analysis presented in Figure \ref{fig 4} already suggests the
importance of radiation force. We will study this problem in more
detail in a future work. If this speculation is correct, the
mechanism of producing outflow in slim disks and hot accretion flows
is quite different. We note in this context that the angular
distribution of the outflow and inflow rates in our simulation are
quite different from the case of hot accretion flows (refer to
Section 3.4). This may be evidence for the different origin of
outflow in slim disk and hot accretion flows.

Combing the cases of slim disks and the hydrodynamical hot accretion
flow, we find that the slope of the radial profile of inflow rate is
quite similar, although the mechanisms of producing outflow in the
two cases are likely different, as we state above. In fact, the
slope is even similar to the case of magnetohydrodynamical hot
accretion flow as well. In that case, the mechanism of producing
outflow is identified to be a Lorentz force such as
magnetocentrifugal force (Yuan et al. 2012; F. Yuan et al. 2013, in
preparation). What is the reason for the same slope in spite of
different mechanisms? Begelman (2012) may provide an answer to the
question of why the hydrodynamical and magnetohydrodynamical hot
accretion flows have the same radial profile of inflow rate, see
also the summary presented in Yuan et al. (2012). Now the similarity
among the three cases seems to indicate that the analysis in
Begelman (2012) also applies to radiation-dominated slim disks.

Finally, our whole investigation presented in this paper is based on
the assumption that the mass accretion rate can be super-Eddington.
However, observations of a large sample of AGNs with 407 sources
show that almost all active galactic nuclei (AGNs) are radiating below $L_{\rm Edd}$
(Kollmeier et al. 2006). Later, Steinhardt \& Elvis (2010) extended
this study to a much larger sample consisting of 62,185 quasars from
the Sloan Digital Sky Survey, and they got a similar conclusion (but
see Kelly \& Shen (2013) for a different opinion). Liu et al. 2013
most recently address this sub-Eddington puzzle. The basic idea is
that because of radiative feedback, the mass accretion rate at the inner
accretion flow can be self-regulated and thus cannot be super-Eddington.
Unfortunately they find that this mechanism cannot fully solve this
problem.

\acknowledgments{We thank Wei-Min Gu and Aleksander Sadowski for
helpful discussions and the anonymous referee for constructive
comments. This work was supported in part by the Natural Science
Foundation of China (grants 11103059, 11121062, 11133005, and
11003052), the National Basic Research Program of China (973 Program
2009CB824800), the CAS/SAFEA International Partnership Program for
Creative Research Teams, and the Fundamental Research Funds for the
Central Universities (No.CQDXWL-2012-019). The simulations were
carried out both at Shanghai Supercomputer Center and the Super
Computing Platform of Shanghai Astronomical Observatory.}

\end{document}